\documentclass[conf]{new-aiaa}
\usepackage[utf8]{inputenc}
\usepackage{graphicx}
\usepackage{amsmath}
\usepackage[version=4]{mhchem}
\usepackage{siunitx}
\usepackage{longtable,tabularx}
\usepackage{textcomp}
\usepackage{float}
\usepackage{lettrine}
\usepackage{hyperref}
\title{SATMO: a Multi-Planet Thermal Analysis Tool for CubeSat Missions}

\author{Alexander G. Chipps \footnote{National Science Foundation Graduate Research Fellow, Department of Aeronautics and Astronautics, 77 Massachusetts Ave., 37-344a, Cambridge, MA 02139, USA}}
\affil{Massachusetts Institute of Technology, Cambridge, Massachusetts, 02139, USA}

\author{Daniel Forgette \footnote{Group Supervisor, Payload Thermal Systems, La Cañada Flintridge, California, 91011, USA}}
\affil{Jet Propulsion Laboratory, La Cañada Flintridge, California, 91011, USA}

\author{Kerri Cahoy \footnote{Professor, Department of Aeronautics and Astronautics, 77 Massachusetts Ave., 37-341, Cambridge, MA 02139, USA, AIAA Associate Fellow}}
\affil{Massachusetts Institute of Technology, Cambridge, Massachusetts, 02139, USA}

\begin{document}

\maketitle

\begin{abstract}
The expansion of commercial launch capabilities has significantly increased opportunities for interplanetary small satellite (SmallSat) missions. As researchers plan for more missions beyond Earth, there is a demand for accessible tools that help better predict and understand the thermal effects on their spacecraft in orbital environments around Earth and other bodies. While commercial thermal analysis tools offer high-fidelity modeling capabilities and results, they are often expensive and require extensive training to be used effectively. This paper details a framework for a user-friendly Satellite Thermal Model (SATMO) to support the early stages of space mission planning for CubeSats orbiting Earth and other Solar System bodies. SATMO is an open-source, MATLAB\textsuperscript{\textregistered}-based, six-node thermal analysis program designed for satellites in low-altitude circular orbits. Although SATMO requires a MATLAB license---typically inexpensive or institutionally provided in academic settings---it remains substantially more accessible than professional thermal analysis software. SATMO requires an internet connection for some features but does not rely on additional MATLAB toolboxes. The SATMO modeling approach is validated with the space industry standard Thermal Desktop\textsuperscript{\textregistered} software, with temperatures comparable to within $\boldsymbol{1.17 ^\circ}$C for a 10 cm $\times$ 10 cm $\times$ 10 cm CubeSat in various configurations, orbiting around primary bodies including Earth, Venus, and Mars. An example use case of SATMO is presented with a Mars-orbiting CubeSat to demonstrate its functionalities and the outputs available to users. SATMO offers increased accessibility to satellite thermal modeling for the research community, enabling quick thermal trade studies and interplanetary mission plans.
\end{abstract}

\section{Nomenclature}

{\renewcommand\arraystretch{1.0}
\noindent\begin{longtable*}{@{}l @{\quad=\quad} l@{}}
$A$ & area, \si[per-mode=symbol]{\m\squared} \\
$a_p$ & albedo factor \\
$\alpha$ & absorptivity \\
$\beta$ & angle between primary body–Sun vector and satellite orbit plane, deg \\
$\beta^*$ & critical beta angle, deg \\
$c$ & constant-pressure specific heat \si[per-mode=symbol]{\joule\per\kg\per\kelvin} \\
$\delta_s$ & declination of the Sun, deg \\
$\Delta t$ & time step, s \\
$\dot{q}$ & heat flux, \si[per-mode=symbol]{\watt\per\m\squared} \\
$\dot{Q}$ & heat rate, W \\
$\varepsilon$ & emissivity \\
$\eta_{panel}$ & solar panel efficiency \\
$f_E$ & eclipse fraction \\
$F$ & view factor \\
$\mathcal{F}$ & scaling factor \\
$G$ & gravitational constant, \num{6.6743e-11}~\si[per-mode=symbol]{\metre\cubed\per\kilogram\per\second\squared} \\
$h$ & orbit altitude, m \\
$in$, $gen$, $out$ & incoming, generated, and outgoing heat indices \\
$i$, $j$ & surface or node indices \\
$i_p$ & equatorial inclination of the primary body, deg \\
$i_{sat}$ & orbital inclination of the satellite, deg \\
$IR$, $sol$, $alb$, $cond$, $htr$, $e$ & infrared, solar, albedo, conduction, heater, and electronic heating indices \\
$J_2$ & J2 orbital perturbation constant \\
$k$ & time index \\
$K$ & conduction coefficient, \si[per-mode=symbol]{\watt\per\kelvin} \\
$L_s$ & primary body-centered longitude of the Sun \\
$m$ & mass, kg \\
$\Omega$ & right ascension of the ascending node, deg \\
$\Omega_s$ & right ascension of the Sun, deg \\
$p$ & primary body index \\
$R$ & radius, m \\
$r$ & distance, m \\
$Sun$, $dark$ & Sun side and dark side of the primary body indices \\
$\sigma$ & Stefan–Boltzmann constant, \num{5.6704e-8}~\si[per-mode=symbol]{\watt\per\metre\squared\per\kelvin\tothe{4}} \\
$T$ & temperature, K \\
$\theta$ & orbital position, deg \\
$\theta_1$ & shadow entry angle, deg \\
$\theta_2$ & shadow exit angle, deg \\
$\vec{v}$ & orbital velocity vector, \si[per-mode=symbol]{\meter\per\s} \\
$\xi$ & solar zenith angle, deg \\
$zen$, $nad$, $+v$, $-v$, $N$, $S$  & denoting zenith, nadir, forward, aft, north, and south directions \\
\end{longtable*}}

\section{Introduction}
\lettrine{I}{nterplanetary} small satellites, or SmallSats, are low-mass spacecraft that resemble traditional satellites but on a smaller scale and that orbit bodies other than Earth. SmallSats are equipped with self-contained power, electronics, instrument payloads, mechanical, avionics, communication, thermal control, attitude control, navigation, and sometimes propulsion systems to explore space, investigate scientific questions, and test new technologies on compact platforms. There is no single, commonly agreed-upon mass-and-size definition for SmallSats; however, this work is consistent with the NASA definition that considers satellites with masses below about 180 kg and volumes less than that of a large kitchen refrigerator ($\approx0.5 - 0.8$ \si[per-mode=symbol]{\meter\cubed}) to fall under the SmallSat class of spacecraft \cite{caldwell_what_2024}. CubeSats are a type of SmallSat in form factor multiples of 1U (10 cm $\times$ 10 cm $\times$ 10 cm) \cite{caldwell_what_2024}. In 2018, alongside the InSight lander, NASA's Jet Propulsion Laboratory (JPL) launched the first pair of interplanetary CubeSats called Mars Cube One (MarCO). MarCO-A and MarCO-B were two 6U spacecraft that deployed from an Atlas V rocket to assist telecommunications for the entry, descent, and landing operations of InSight \cite{klesh_marco_2015, schoolcraft_marco_2017}. With the success of MarCO, other SmallSat missions have been planned for Mars. For example, the Escape, Plasma, and Acceleration Dynamics Explorers (ESCAPADE) mission aims to study the solar wind and Martian magnetosphere interactions via twin SmallSats, each with a dry mass of about 201 kg \cite{parker_escapade_2022}. Among 10 interplanetary CubeSats that traveled to the Moon with Artemis I were Lunar IceCube \cite{malphrus_lunar_2019} and Lunar Flashlight \cite{cohen_lunar_2020}, two 6U cislunar spacecraft, each used for prospecting and studying water ice deposits and other volatile compounds on the lunar surface. Venus is also a future target for interplanetary CubeSat missions; some researchers suggest Venus as a destination for demonstrating aerocapture with SmallSats \cite{malphrus_interplanetary_2021}. Mission study concepts for Venus include Cupid's Arrow \cite{sotin_cupids_2018} and CubeSat UV Experiment (CUVE) \cite{cottini_cuve_2018}, SmallSats for respectively studying the noble gases and a mysterious UV absorbing agent in Venus' atmosphere.

Interplanetary SmallSats must overcome greater challenges in propulsion systems (such as extremely demanding total $\mathrm{\Delta V}$ requirements), long-distance telecommunications, and radiation tolerance when compared with Earth-orbiting spacecraft \cite{malphrus_interplanetary_2021}. Thermal control is another crucial aspect of interplanetary SmallSat design, since these spacecraft and their payloads must function in thermal environments that can be more extreme than those in Earth orbit, while still complying with the SmallSat form factor. As more interplanetary satellite missions are planned, there is greater demand for tools that help mission designers and engineers better understand the general orbital thermal environment that a spacecraft will experience during operation. Commercial thermal modeling tools like Thermal Desktop (TD), while offering high fidelity, are complex, expensive, and time-consuming to learn. Other simple and low-cost tools offer quick ways to assess thermal management strategies in various orbital configurations prior to more complex modeling. Programs like the Adaptive Thermal Modeling Tool (ATMT) \cite{richmond_adaptive_2010} and SatTherm \cite{vanoutryve_thermal_2008, allison_sattherm_2009} are multi-node thermal analysis tools based in MATLAB that are designated for studying the transient temperature behavior of small spacecraft. ATMT and SatTherm include the thermal effects of environmental heating rates, conduction between adjacent satellite surfaces, and internal radiation between surfaces with user-specified properties. Users of these programs have the ability to model satellites in circular or elliptical orbits around bodies other than Earth, but ATMT and SatTherm have only been demonstrated and validated against TD for Earth-orbiting SmallSats in the literature to temperatures within $5^\circ$C \cite{richmond_adaptive_2010} and $4^\circ$C \cite{vanoutryve_thermal_2008, allison_sattherm_2009}, respectively. Versteeg et al.~\cite{versteeg_preliminary_2018} also demonstrates a six-node thermal code for Earth-orbiting satellites, with its analysis strongly dependent on beta angle, $\beta$, the angle between the primary body-Sun vector and the satellite orbital plane, since this angle significantly influences orbital heating rates \cite{versteeg_preliminary_2018}. The approach taken in \cite{versteeg_preliminary_2018} is rooted in work by \cite{gilmore_spacecraft_2002} (and seemingly \cite{rickman_introduction_2014}), but has not been explicitly validated. Even simpler approaches exist which represent satellites using single-node lumped capacitance models such as those in \cite{totani_one_2013} and \cite{anh_thermal_2016} for extremely rapid screening of orbital thermal environments.

While robust in their modeling features, previous non-commercial or open source thermal analysis tools have not placed heavy emphasis on multi-planet thermal modeling. To further support interplanetary space missions (e.g., for planetary climate studies, demonstrations of long-distance communications, etc.), this work presents the Satellite Thermal Model (SATMO). SATMO is an open-source, MATLAB-based, six-node thermal model for CubeSats in circular orbits around Earth and beyond, which builds upon the features of previous orbital thermal analysis tools and approaches such as \cite{richmond_adaptive_2010, vanoutryve_thermal_2008,versteeg_preliminary_2018,gilmore_spacecraft_2002,rickman_introduction_2014,rickman_simplified_2002}. The SATMO modeling approach is validated using TD and demonstrated with a 1U CubeSat in various configurations, orbiting around primary bodies including Earth, Venus, and Mars. Since many CubeSats are designed by research and student groups, a major motivation for this work and its predecessors is to empower anyone on these teams to understand the thermal environment that their spacecraft will experience, without extensive funding or expertise in thermal engineering. SATMO is available for download at \url{https://github.com/alexchipps/SATMO}, and it is fully supported in MATLAB R2025b. Users may find that SATMO runs successfully in other MATLAB versions, although compatibility has not been exhaustively tested. Ultimately, SATMO offers increased accessibility to planetary satellite thermal modeling for the research community, enabling research groups to quickly perform thermal trade studies and plan interplanetary CubeSat missions.

\section{Thermal Model Development}
\label{sec: Thermal Model Development}
\subsection{General SATMO Architecture}
\begin{figure}[H]
\centering
\includegraphics[scale=0.45]{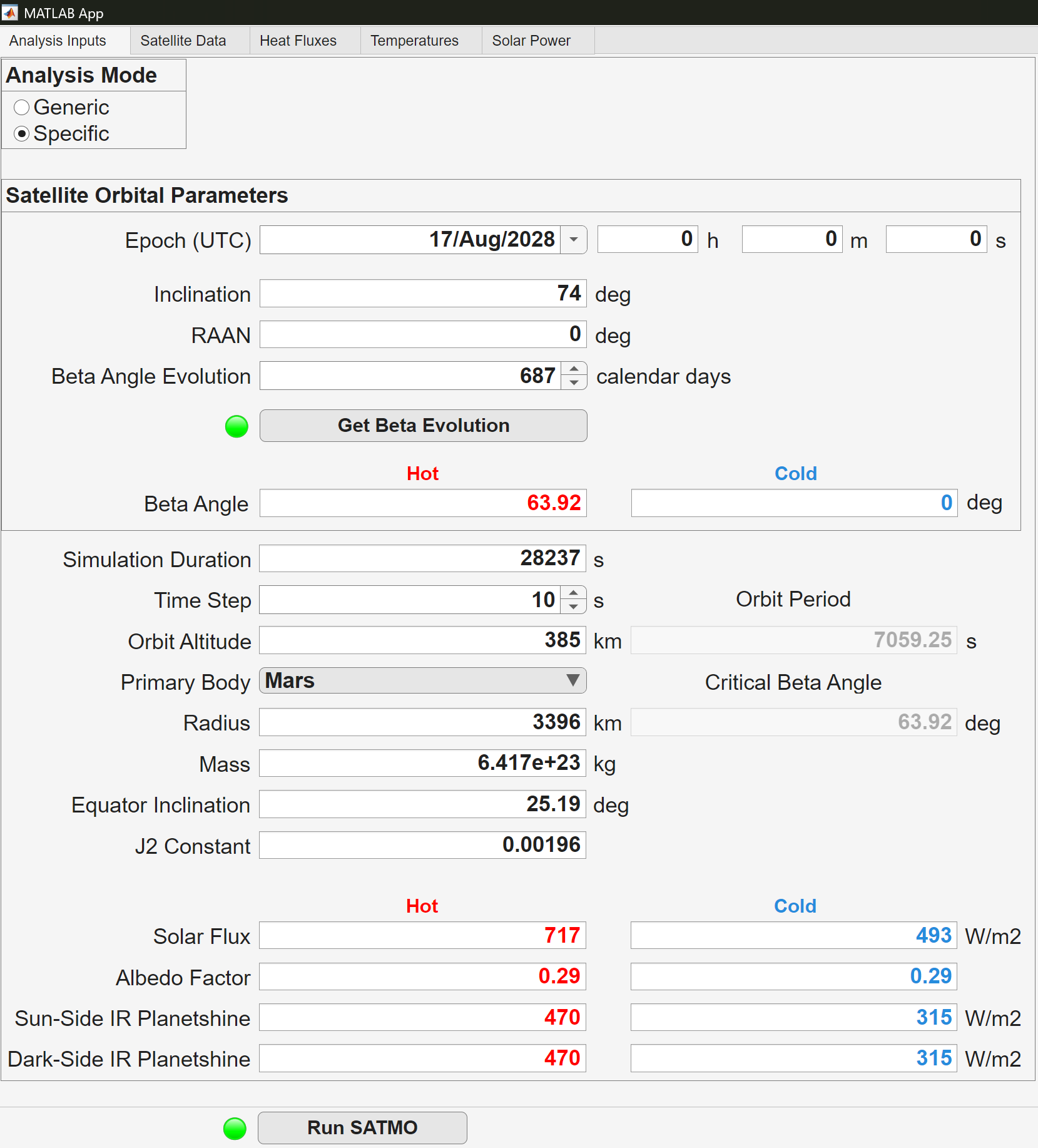}
\caption{SATMO Analysis Inputs tab with example user inputs for the specific analysis mode.}
\label{fig: SATMO UI Analysis Inputs} 
\end{figure}

\begin{figure}[H]
\centering
\includegraphics[scale=0.45]{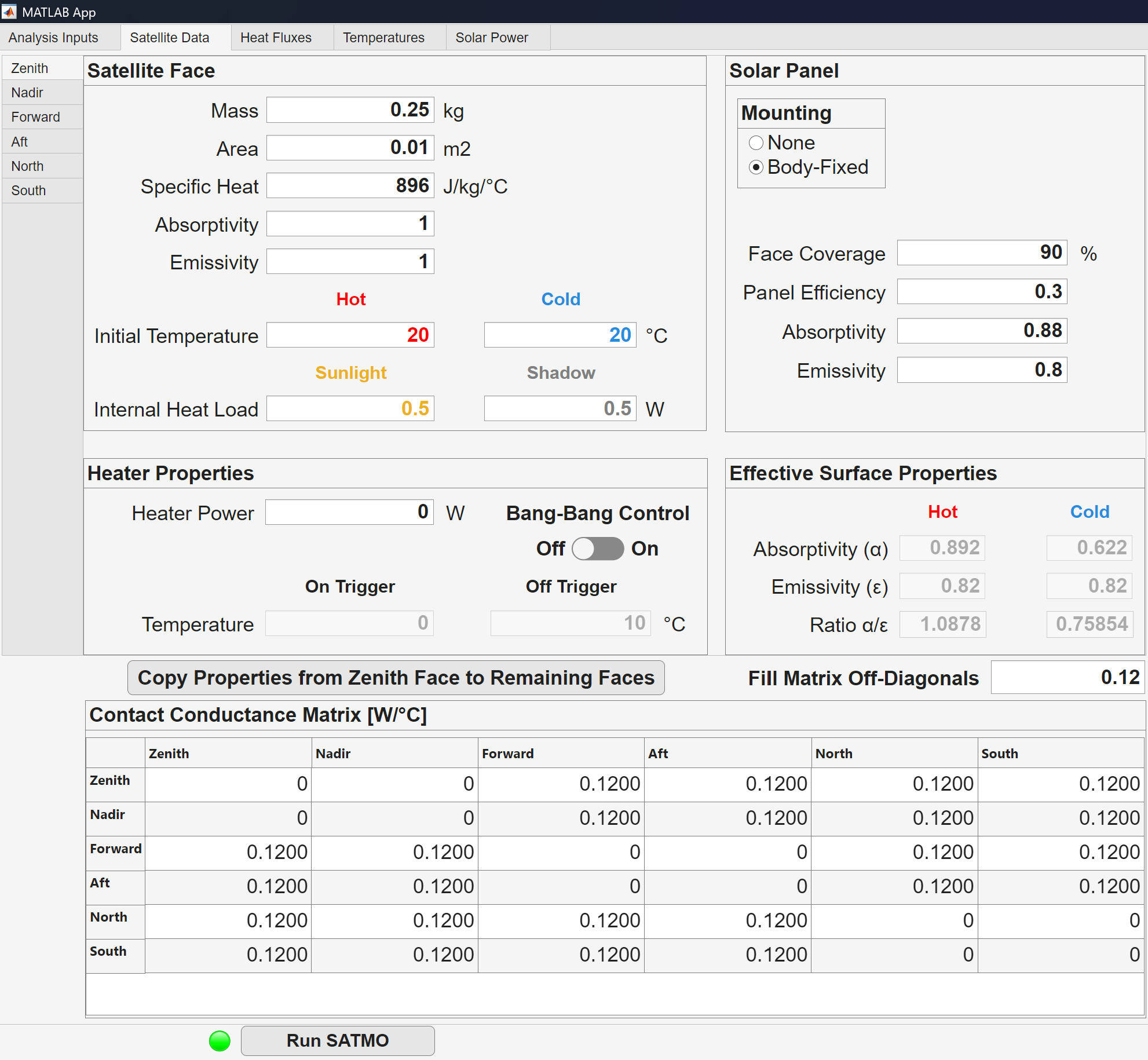}
\caption{SATMO Satellite Data tab with example user inputs for the Zenith face. Each satellite face has inputs of this form.}
\label{fig: SATMO UI Satellite Data} 
\end{figure}

SATMO offers a straightforward approach to satellite orbital thermal modeling and analysis---users simply specify the characteristics of their mission, satellite, and orbital environment within two main input tabs of a MATLAB application, as shown in Fig.~\ref{fig: SATMO UI Analysis Inputs} and Fig.~\ref{fig: SATMO UI Satellite Data}. SATMO thermal simulations are performed on a satellite represented as a six-sided box with one face-centered node per surface in a circular orbit about a primary body. After the thermal simulations are complete, net absorbed heat fluxes, satellite surface temperatures, and solar panel power outputs are populated in three respective output tabs. In essence, SATMO reads in the user inputs, and for each beta angle specified, the program iteratively calculates the environmental heat fluxes on each satellite surface and updates the surface's temperature over user-specified time steps using an energy balance technique about each face-centered node. The SATMO app and its supporting files are available at \url{https://github.com/alexchipps/SATMO} and can be accessed and modified by anyone.

On the Analysis Inputs tab (Fig.~\ref{fig: SATMO UI Analysis Inputs}), users must specify the simulation duration and time step, satellite orbit altitude, primary body (body orbited by the satellite), as well as the primary body's radius, mass, equatorial inclination with respect to the body's orbital plane around the Sun, J2 constant, and constant values used to calculate the environmental heat fluxes incident upon the satellite surfaces. These constants include the hot and cold case values for solar fluxes at the primary body's distance from the Sun, albedo factors (i.e., the fraction of incident solar energy that is reflected off of the primary body), and thermal infrared (IR) fluxes (i.e., IR planetshine of the primary body). Users must also choose between two types of thermal analysis modes, \textbf{generic} or \textbf{specific}.

In the generic analysis mode, hot and cold case thermal simulations are performed over a range of discrete beta angles selected by the user. In both cases, the thermal simulations are performed at each beta angle with the upper and lower bounds of the environmental heating parameters and effective surface properties, respectively. Standard industry practice is to sweep across beta angles in increments of $\approx 5-10^\circ$ to determine the worst-case thermal conditions for each relevant component in the satellite, which is highly dependent on spacecraft geometry and configuration.

In the specific analysis mode, the user selects one hot-case and one cold-case beta angle at which the thermal simulations will be performed. In this mode, SATMO also provides the evolution of $\beta$ and percentage of orbit in sunlight over calendar date, once additional mission parameters---such as simulation start date and time, spacecraft orbital inclination, and right ascension of the ascending node (RAAN)---are specified. The hot-case simulation applies the upper bounds of the environmental heating parameters and effective surface properties, whereas the cold-case simulation applies the corresponding lower bounds. Currently, the specific analysis option exists for the Moon, Pluto, and all major planets in the Solar System, but this work only evaluates SATMO for Venus-, Earth-, and Mars-orbiting satellites. For each thermal simulation conducted at a specified beta angle, the Sun-primary body system is treated as fixed in space, and only the satellite is assumed to move along its orbit.

On the Satellite Data tab (Fig.~\ref{fig: SATMO UI Satellite Data}), the user inputs each satellite face's constant mass, area, specific heat, absorptivity, emissivity, initial temperature, internal heat loads from electronics, conductance coefficient between adjacent surfaces, and optional parameters for heater power and controls. Body-mounted solar panels are other optional inputs, for which the user must specify the solar panel area coverage, efficiency, and optical properties.

\subsection{Parameters Influencing the Orbital Thermal Environment}
\begin{figure}[H]
\centering
\includegraphics[scale=1.135]{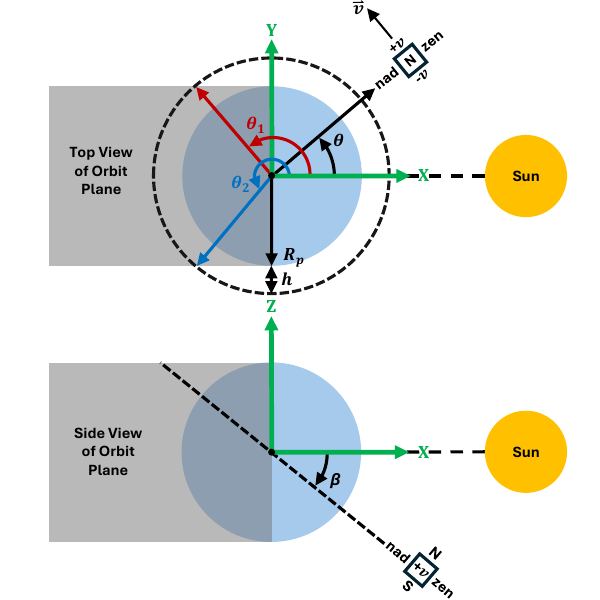}
\caption{Depiction of an orbiting box and nomenclature used in SATMO, adapted from \cite{rickman_introduction_2014}. Figure not to scale.}
\label{fig: Orbital Configuration} 
\end{figure}

Figure \ref{fig: Orbital Configuration} is a representative configuration for SATMO's orbital thermal analysis. This figure displays a six-sided orbiting box using nadir-pointing attitude control with surfaces defined in Table \ref{tab: SATMO surface nomenclature}. The box is orbiting at an altitude $h$ above the surface of its primary body with radius $R_p$, and it is positioned at an angle $\theta$ measured from solar noon (i.e., the subsolar point of the orbit). Using a cylindrical umbral shadow approximation, the satellite is in the eclipse (shadow) zone of its orbit (i.e., on the night side of its primary body) when the orbit angle $\theta$ is between the shadow entry angle $\theta_1$ and shadow exit angle $\theta_2$. In reality, the shadow zone is conical with umbral and penumbral regions. However, it is reasonable to assume a cylindrical shadow zone while neglecting penumbral regions in the thermal model for spacecraft in low-altitude (i.e., $h \ll R_p$) orbits \cite{rickman_simplified_2002}. Relative to the Sun, the orbit plane is oriented by $+\beta$ as demonstrated in the bottom graphic of Fig.~\ref{fig: Orbital Configuration}. $\beta \in \left[-90^\circ, 90 ^\circ \right ]$ is conventionally negative when the Sun is south of the orbit plane and positive when the Sun is north of the orbit plane. Alternatively, \cite{gilmore_spacecraft_2002} defines $\beta$ as, ``... positive if the satellite appears to be going counterclockwise around the orbit as seen from the Sun, negative if clockwise.'' The mathematical definition of $\beta$ is shown in Eq.~\eqref{eqn: betaAngle} from \cite{gilmore_spacecraft_2002}\footnote{All trigonometric functions throughout this work are evaluated in degrees unless specified otherwise.}.

\begin{equation}
\label{eqn: betaAngle}
\beta = \sin^{-1} \left( \cos(\delta_s) \sin(i_{sat}) \sin(\Omega - \Omega_s) + \sin(\delta_s) \cos(i_{sat}) \right)
\end{equation}
Here, the solar declination $\delta_s$ is measured relative to the primary body's equatorial plane, positive towards the north pole. The orbital inclination of the satellite $i_{sat}$ is the angle between the satellite's orbit plane and the primary body's equatorial plane. The satellite orbit's RAAN and the right ascension (RA) of the Sun are denoted by $\Omega$ and $\Omega_s$, respectively. According to the TD user manual \cite{panczak_ansys_2024}, RA is defined as the angle measured in the equatorial plane from the vernal equinox, increasing counter-clockwise when viewed from the north pole of the primary body (using the right-hand rule). The ascending node refers to the point where the satellite's orbit intersects the equatorial plane from south to north \cite{panczak_ansys_2024}.

\begin{table}[H]
\caption{\label{tab: SATMO surface nomenclature} Surface nomenclature for a satellite represented as a six-sided orbiting box with nadir-pointing attitude control in SATMO.}
\centering
\begin{tabular}{l c p{2in}}
\hline
\textbf{Surface name} & \textbf{Symbol} & \textbf{Description} \\
\hline
Zenith & $zen$ & Faces away from the primary body \\
Nadir & $nad$ & Faces the primary body, tangent to its surface \\
Forward & $+v$ & Faces in the direction of the orbital velocity vector \\
Aft & $-v$ & Faces in the opposite direction of the orbital velocity vector \\
North & $N$ & Faces in the out-of-page direction when viewing the orbit plane from the top down \\
South & $S$ & Faces in the into-the-page direction when viewing the orbit plane from the top down \\
\hline
\end{tabular}
\end{table}

The fraction of a given orbit in eclipse decreases as $\lvert\beta\rvert$ increases. In modeling eclipse conditions, SATMO assumes that shadowing arises solely from sunlight obstruction by the primary body, neglecting shadowing contributions from planetary ring structures and moons when present. As noted in \cite{gilmore_spacecraft_2002}, there will be some critical beta angle, $\beta^* \in \left[0^\circ, 90 ^\circ \right ]$, for which any $\lvert \beta \rvert > \beta^*$ will expose a spacecraft's orbit to constant sunlight. Equation (\ref{eqn: betaCrit}) provides a critical beta angle estimate for low circular orbits \cite{gilmore_spacecraft_2002}.

\begin{equation}
\label{eqn: betaCrit}
\beta^* = \sin^{-1} \left( \frac{R_p}{R_p + h} \right)
\end{equation}

\begin{figure}[H]
\centering
\includegraphics[width=3.75in, height=2.25in]{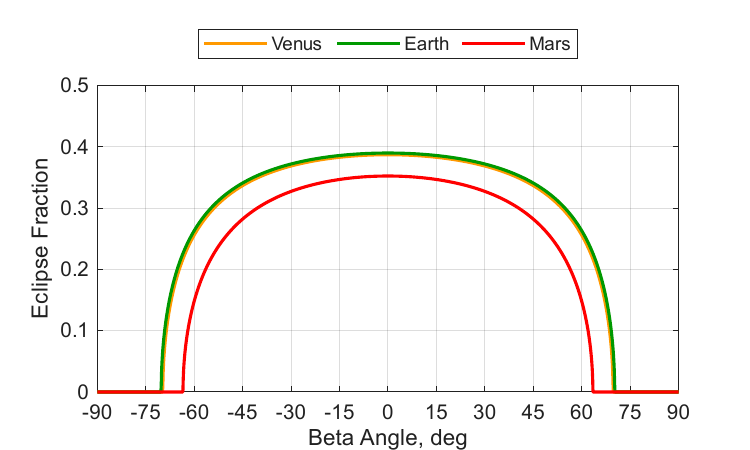}
\caption{Eclipse fraction versus beta angle for a spacecraft in 400-km circular orbit around Venus, Earth, and Mars. The lines for Venus and Earth are nearly overlapping due to their similar sizes.}
\label{fig: Eclipse Fraction} 
\end{figure}

The fraction of an orbit eclipsed by the primary body in relation to sunlight is also provided by \cite{gilmore_spacecraft_2002} in Eq.~\eqref{eqn: f_E}.

\begin{equation}
\label{eqn: f_E}
f_E =
\begin{cases}
\frac{1}{180^\circ} \cos^{-1}\left(\frac{\sqrt{h^2 + 2R_ph}}{\left(R_p + h\right) \cos\left(\beta\right)}\right), & \lvert \beta \rvert < \beta^*\\
0, & \lvert \beta \rvert \geq \beta^*\\
\end{cases}
\end{equation}
Figure \ref{fig: Eclipse Fraction} displays $f_E$ as a function of $\beta$ for Venus-, Earth-, and Mars-orbiting satellites at $h = 400$ km, where the lines for Earth and Venus are nearly overlapping due to their similar sizes. In cases where $f_E > 0$, it is important to know when the spacecraft will pass through its primary body's shadow region. Equations \eqref{eqn: shadowEntry} and~\eqref{eqn: shadowExit} present the shadow entry and exit angles, respectively, measured from orbit noon and derived by \cite{rickman_introduction_2014}.

\begin{equation}
\label{eqn: shadowEntry}
\theta_1 =
\begin{cases}
180^\circ - \sin^{-1}\left( \sqrt{\cos^{-2}(\beta) \left[ \left( \frac{R_p}{R_p + h} \right)^2 - \sin^{2}(\beta)\right]} \right), & f_E > 0\\
180^\circ, & f_E = 0\\
\end{cases}
\end{equation}

\begin{equation}
\label{eqn: shadowExit}
\theta_2 =
\begin{cases}
360^\circ - \theta_1, & f_E > 0\\
180^\circ, & f_E = 0\\
\end{cases}
\end{equation}
For the degenerate case where the satellite never passes into eclipse, $\theta_1$ and $\theta_2$ become intermediate parameters that are set to $180^\circ$. This allows the relations to hold in Eq.~\eqref{eqn: F_sol} and Eq.~\eqref{eqn: F_alb}, which are introduced in Section \ref{sec: View Factors and Scaling Factors}.

The combination of the parameters $\delta_s$, $i_{sat}$, $\Omega$, and $\Omega_s$ in Eq.~\eqref{eqn: betaAngle} allow the spacecraft orbital thermal environment to simply be analyzed with a discrete $\beta$ approach. However, space mission designers may want to know the approximate time and duration of different $\beta$ scenarios, which can be determined if $\beta$ is known as a function of time. Therefore, this work outlines a method for predicting $\beta$ as a function of calendar date, based on changes in $\delta_s$, $\Omega$, and $\Omega_s$ over time.

Beginning with $\Omega$, a primary body's equatorial bulge and uneven mass distribution cause the satellite orbit's RAAN to shift at a rate of $\dot\Omega$ \si[per-mode=symbol]{deg\per\s} via the J2 effect, which is the only orbital perturbation included in SATMO. All other perturbations (i.e., atmospheric drag, solar radiation pressure, and third-body influences) are neglected, and the satellite's motion is otherwise modeled using two-body dynamics. As a result, the satellite's orbital period remains constant. $\dot\Omega$ can be calculated for a circular orbit using Eq.~\eqref{eqn: RAANdot} from \cite{rickman_introduction_2014}.

\begin{equation}
\label{eqn: RAANdot}
\dot\Omega =
\begin{cases}
- \frac{3}{2} J_2 \left( \left( {\frac{R_p}{R_p + h}} \right)^2 \cos(i_{sat}) \sqrt{\frac{Gm_{p}}{(R_p + h)^3}}  \right), & 0^\circ < i_{sat} < 180^\circ \\
0, & i_{sat} = 0^\circ \text{ or } 180^\circ
\end{cases}
\end{equation}
Equation \eqref{eqn: RAANdot} results in a westward precessing ascending node when $0^\circ < i_{sat} < 90^\circ$, an eastward precessing ascending node when $90^\circ < i_{sat} < 180^\circ$, and no nodal precession ($\dot\Omega = 0$ \si[per-mode=symbol]{\degree\per\s}) for a zero-inclination orbit. $\Omega$ is updated after a time step $\Delta t$ according to Eq.~\eqref{eqn: OmegaPlus1}.

\begin{equation}
\label{eqn: OmegaPlus1}
\Omega_{k+1} = \Omega_{k} + \dot\Omega\Delta t
\end{equation}

Next, $\Omega_s$ is calculated in Eq.~\eqref{eqn: OmegaS} per \cite{meeus_astronomical_1998},

\begin{equation}
\label{eqn: OmegaS}
\Omega_s = \operatorname{mod} \Bigl(\operatorname{atan2} \bigl( \cos(i_p)\sin(L_s), \cos(L_s) \bigr), 360^\circ \Bigr)\end{equation}
where $i_p$ is the angle measured from the primary body's orbital plane around the Sun to its equatorial plane. $L_s$ ranges from $0^\circ$ to $360^\circ$ along the planet's orbital plane. SATMO retrieves apparent solar longitude values via internet connection from JPL's Horizons System API \cite{jpl_solar_system_dynamics_group_horizons_2025} over a desired period, outputting $L_s$ with respect to various planets every 6 hours. JPL defines the value as, ``The apparent target-centered longitude of the Sun (`apparent L\textunderscore s') as seen at the target when the light recorded by the observer at print-time reflected off the target. It is referred to a coordinate system where the x-axis is the equinox direction defined by the target's instantaneous IAU2015 pole direction and heliocentric orbit plane at reflection time, and is measured positively in an eastward direction (counter-clockwise around the positive pole of the Solar System angular momentum vector)'' \cite{jpl_solar_system_dynamics_group_horizons_2025}. IAU2015 refers to standards for planetary cartographic coordinates and rotational elements defined by the International Astronomical Union in their 2015 report \cite{archinal_report_2018}. SATMO assumes that the light-time delay resulting in the apparent perceived solar longitude is negligible for this scale of thermal analysis precision, as the delay is on the order of 10 minutes from the Sun to Mars in the worst case. SATMO approximates the raw apparent solar longitude value from Horizons as $L_s$. 

Lastly, $\delta_s$ is calculated using TD's formulation in Eq.~\eqref{eqn: solarDeclination}. Beta angle evolution can be determined as long as the calendar dependence on $L_s$ can be found. Refer to \cite{jpl_solar_system_dynamics_group_horizons_2025} to see the time spans for which ephemeris data exist for various primary bodies.

\begin{equation}
\label{eqn: solarDeclination}
\delta_s = i_p\sin(L_s)
\end{equation}

\subsection{Energy Balance and Heating Contributions}
\label{sec: Energy Balance}
SATMO uses the energy balance technique  to calculate the time-dependent temperature change of a given satellite surface, $i$ in Eq.~\eqref{eqn: qNet}.

\begin{equation}
\label{eqn: qNet}
\dot{Q}_{{in}_i} + \dot{Q}_{{gen}_i} - \dot{Q}_{{out}_i} = m_ic_i\frac{\partial T_i}{\partial t}
\end{equation}

The primary incoming environmental heating contributions onto a satellite surface are due to IR radiation emitted from the primary body, direct solar radiation, and reflected solar radiation off the primary body (albedo). Equations~\eqref{eqn: qIR}--\eqref{eqn: qAlb} present the environmental heating contributions for an individual surface. Spacecraft collisions with individual molecules in the outermost parts of planetary atmospheres and collisions with charged particles in places like Earth's Van Allen radiation belts yield additional heating sources called free molecular heating (FMH) and charged-particle heating (CPH), respectively \cite{gilmore_spacecraft_2002, rickman_introduction_2014}. However, FMH is typically only encountered  after payload fairing and booster separation during launch ascent, and CPH is generally only significant for cryogenic system thermal analysis \cite{gilmore_spacecraft_2002}. Thus, SATMO neglects FMH and CPH effects in its orbital heat load model.

\begin{equation}
\label{eqn: qIR}
\dot{Q}_{{IR}_i} =  \varepsilon_i A_i F_i \dot{q}_{IR}
\end{equation}

\begin{equation}
\label{eqn: qSol}
\dot{Q}_{{sol}_i} =  \alpha_i A_i \mathcal{F}_{sol \to i} \dot{q}_{sol}
\end{equation}

\begin{equation}
\label{eqn: qAlb}
\dot{Q}_{{alb}_i} =  a_p \alpha_i A_i F_i \mathcal{F}_{alb \to i} \dot{q}_{sol}
\end{equation}

Here, $\varepsilon$ and $\alpha$ are the thermo-optical properties of a surface with irradiated or radiating surface area, $A$. $\varepsilon$ is linked to radiation in thermal IR wavelengths (2.5 \si[per-mode=symbol]{\micro\meter} to 50 \si[per-mode=symbol]{\micro\meter}), while $\alpha$ is linked to radiation in solar wavelengths (240 nm to 2.5 \si[per-mode=symbol]{\micro\meter}) \cite{cech_young_next-generation_2023}. Various combinations of $\alpha$ and $\varepsilon$ can be input by SATMO users to test the thermal response of their spacecraft to different surfaces and coatings. When solar panels are mounted on a satellite surface, SATMO computes the effective surface absorptivity and emissivity using area-weighted averages of the constituent materials. Depending on the selected analysis mode, the program also accounts for solar panel efficiency in the calculation for effective absorptivity. For hot-case bounding analyses, SATMO assumes all solar energy incident on a solar panel is absorbed as heat and none of it is converted into electrical energy. For cold-case analyses, the model conservatively assumes that power is continuously drawn from the solar panels when they are illuminated, meaning a fraction $\eta_{panel}$ of the incident solar energy is converted to electricity without contributing to heating.

As an example, consider a surface that is covered 75\% by a solar panel (with $\alpha_{panel} = 0.90$, $\varepsilon_{panel} = 0.80$, and $\eta_{panel} = 0.30$) and 25\% by white paint (with $\alpha_{paint} = 0.10$ and $\varepsilon_{paint} = 0.90$). The effective thermo-optical properties for the hot and cold cases are then found in Eq.~\eqref{eqn: aHot}--\eqref{eqn: eCold}.

\begin{equation}
\label{eqn: aHot}
\alpha_{hot} = 0.75\alpha_{panel} + 0.25\alpha_{paint} = 0.700
\end{equation}

\begin{equation}
\label{eqn: eHot}
\varepsilon_{hot} = 0.75\varepsilon_{panel} + 0.25\varepsilon_{paint} = 0.825
\end{equation}

\begin{equation}
\label{eqn: aCold}
\alpha_{cold} = 0.75(\alpha_{panel} - \eta_{panel}) + 0.25\alpha_{paint} = 0.475
\end{equation}

\begin{equation}
\label{eqn: eCold}
\varepsilon_{cold} = \varepsilon_{hot} = 0.825
\end{equation}
While the intrinsic material properties of the constituents do not change, Eqs.~\eqref{eqn: aHot}--\eqref{eqn: eCold} are convenient methods to reflect how a surface as a whole interacts thermally with its environment.

The factors $F_i$, $\mathcal{F}_{{sol}{\to}i}$, and $\mathcal{F}_{{alb}{\to}{i}}$ in Eqs.~\eqref{eqn: qIR}--\eqref{eqn: qAlb} adjust the intensity of the incoming heat loads depending on a surface's orientation to and/or distance from the heating sources. The local view factor $F_i$ to the primary body scales the intensity of the incoming albedo and IR heating contributions based on the surface's orientation relative to and altitude above the primary body. The scaling factors $\mathcal{F}_{{sol}{\to}i}$ and $\mathcal{F}_{{alb}{\to}{i}}$ scale the solar and albedo heating intensity based on the satellite's geometry and angle from the subsolar point of the orbit. These factors are described further in Section \ref{sec: View Factors and Scaling Factors}.

The IR flux $\dot{q}_{IR}$ emitted by the primary body into space and the solar flux $\dot{q}_{sol}$ at the primary body's distance from the Sun vary across surface locations, seasons, and distances from the Sun. Planetary albedo factor $a_p$---often shortened to `albedo'---refers to the fraction of incident solar energy reflected off a body, and it, too, may vary spatially and temporally. However, SATMO assumes that all surfaces are diffuse and that the user input values for $\dot{q}_{IR}$, $\dot{q}_{sol}$, and $a_p$ are constant. For some bodies, like Mercury and the Moon, IR planetshine is vastly different on the Sun side of the body compared to the dark side \cite{gilmore_spacecraft_2002}. Therefore, SATMO offers the ability for users to enter distinct, constant values for $\dot{q}_{IR}$ on each side of the terminator following Eq.~\eqref{eqn: qIR day/night}.

\begin{equation}
\label{eqn: qIR day/night}
\dot{q}_{IR} =
\begin{cases}
\dot{q}_{IR_{Sun}}, & \xi \leq 90^\circ  \text{ or } \xi \geq 270^\circ \\
\dot{q}_{IR_{dark}}, & \text{otherwise}
\end{cases}
\end{equation}
where $\xi$ is a measure of the angular distance between the planet-spacecraft vector and the solar vector. Mathematically, $\cos({\xi}) = \cos({\theta})\cos({\beta})$ for $\theta, \beta \in \left[-90^\circ,90^\circ\right]$. Alternatively, simple approximations for IR planetshine can utilize the effective blackbody surface temperature of the primary body using Eq.~\eqref{eqn: qIR s-b law} per the generalized Stefan-Boltzmann Law.
\begin{equation}
\label{eqn: qIR s-b law}
\dot{q}_{\mathrm{IR}} =
\begin{cases}
\varepsilon_p \, \sigma \, T_{p,\mathrm{Sun}}^4, & \xi \leq 90^\circ \ \text{or} \ \xi \geq 270^\circ, \\
\varepsilon_p \, \sigma \, T_{p,\mathrm{dark}}^4, & \text{otherwise.}
\end{cases}
\end{equation}

The solar flux as a function of any primary body's center-to-center distance from the Sun is given by Eq.~\eqref{eqn: qSolDist}. This equation is derived from dividing the Sun's total power output by the surface area of a reference sphere centered at the Sun with radius $r$, resulting in the inverse-square relationship between solar flux and the primary body's distance from the Sun.
\begin{equation}
\label{eqn: qSolDist}
\dot{q}_{sol}(r) = \varepsilon_{Sun}\sigma T_{Sun}^4 \left (\frac{R_{Sun}}{r} \right )^2
\end{equation}

In addition to environmental heating contributions, SATMO also calculates conduction between adjacent satellite surfaces using Eq.~\eqref{eqn: qCond}.

\begin{equation}
\label{eqn: qCond}
\dot{Q}_{{cond}_i} =  \sum\limits^{6}_{j=1}K_{i \to j} \left( T_j - T_i \right)
\end{equation}
The coefficient $K$ in Eq.~\eqref{eqn: qCond} represents the thermal conductance between surfaces in direct physical contact. Conduction can lead to heating or cooling, depending on the direction of heat transfer between a given surface and its neighboring surfaces. For instance, conduction occurs between the Zenith surface and the Forward, Aft, North, and South surfaces when a temperature difference exists. Additional conductive paths may occur between non-adjacent faces if an internal component thermally connects them. SATMO users must populate a conduction coefficient matrix with their inputs prior to performing the thermal simulation. An example of such a matrix is presented in Section \ref{sec: SATMO Validation} of this work. Contact conductance depends largely on the contact area of the joint, with additional influences from surface roughness, material properties, and interface method (press-fit, bolted, glued, etc.) \cite{richmond_adaptive_2010}. References are provided in \cite{gilmore_spacecraft_2002} that can be used to approximate conduction coefficient values for various configurations of contacting surfaces.

The outgoing heat transfer rate term, $\dot{Q}_{out}$ in Eq.~\eqref{eqn: qOut}, represents the net radiation exchange between a satellite surface and space itself.
\begin{equation}
\label{eqn: qOut}
\dot{Q}_{{out}_i} = \varepsilon_i A_i \sigma (T_i^4 - T_{space}^4)
\end{equation}
where $T_{space}$ is the temperature of deep space, approximated as 2.73 \si[per-mode=symbol]{\K}, following TD and \cite{richmond_adaptive_2010}. Also, it is assumed the spacecraft's view to space is not obstructed by any objects, including the primary body and its moons, during spacecraft-deep space radiation exchange. Like \cite{versteeg_preliminary_2018}, SATMO simplifies the thermal simulation by neglecting internal radiation exchange between surfaces.

Finally, the heat generation term, $\dot{Q}_{gen}$ in Eq.~\eqref{eqn: qGen}, consists of the power from a heater designated for the respective surface as well as a heating contribution from other internal satellite electronics sources like avionics and instruments.

\begin{equation}
\label{eqn: qGen}
\dot{Q}_{{gen}_i} = \dot{Q}_{{htr}_i} + \dot{Q}_{{e}_i}
\end{equation}

Following \cite{versteeg_preliminary_2018}, a discrete integration approach is applied to Eq.~\eqref{eqn: qNet} in which $\frac{\partial T}{\partial t}$ is approximated as $\frac{\Delta T}{\Delta t}$ to advance the temperature $T$ of a surface over small time steps. The updated temperature for a surface at time k + 1 is presented in Eq.~\eqref{eqn: updatedT}.

\begin{equation}
\label{eqn: updatedT}
T_{i,k+1} = T_{i,k} + \frac{\Delta t}{m_ic_i}\left( \dot{Q}_{{IR}_{i,k}} + \dot{Q}_{{sol}_{i,k}} + \dot{Q}_{{alb}_{i,k}} + \dot{Q}_{{cond}_{i,k}} + \dot{Q}_{{gen}_{i,k}} - \dot{Q}_{{out}_{i,k}} \right)
\end{equation}

Some academic and industry satellite thermal models utilize higher order numerical integration schemes \cite{garzon_thermal_2018, krishnaprakas_comparison_1998, vanoutryve_thermal_2008} or linearize the terms in Eq.~\eqref{eqn: qNet} \cite{anh_thermal_2016} to advance the temperature solution, but this work demonstrates that a discrete integration approach over small time steps (e.g., $\Delta t = 1.0$ s used in Section \ref{sec: SATMO Validation}) provides sufficient accuracy in results compared to TD's finite difference methods. However, SATMO users may need to reduce their simulation time step if their input satellite characteristics cause the algorithm to diverge.

\subsection{View Factors and Scaling Factors}
\label{sec: View Factors and Scaling Factors}
In this section, view factors are introduced to scale the intensity of the incoming heat fluxes on each surface with their orientation relative to the primary body, in accordance with the geometry in Fig.~\ref{fig: Orbital Configuration} and formulae adapted from \cite{rickman_introduction_2014}. These factors assume a nominal attitude control scenario of a nadir-pointing satellite. First, the local view factor to the primary body, relevant for the albedo and IR heating contributions, is expressed in Eq.~\eqref{eqn: F_local}. 
\begin{equation}
\label{eqn: F_local}
F_{i} =
\begin{cases} 
    0, & i = zen \\

\frac{1}{2} -
\frac{1}{180^\circ}\,\sin^{-1}\!\left(
\sqrt{1 - \left(\frac{R_p}{R_p+h}\right)^2} \right) - \frac{2}{\pi}\,
\sin\!\left[ 2\,\sin^{-1}\!\left( \sqrt{1 - \left(\frac{R_p}{R_p+h}\right)^2} \right) \right], & i = nad \\
    
    \left(\frac{R_p}{R_p + h}\right)^2, & i = +v, -v, N, S
\end{cases}
\end{equation}
where the Zenith surface is geometrically blind to the orbiting planet at all times, the Nadir surface is tangent to the surface of the primary body below, and the remaining surfaces are perpendicular to the Nadir surface. The formulations in Eq.~\eqref{eqn: F_local} assume the spacecraft is orbiting at a low enough altitude (i.e., $h \ll R_p$) so that the entire field of view of any surface is uniformly illuminated \cite{rickman_introduction_2014, rickman_simplified_2002}. This uniform illumination assumption is a simplification, and methods for handling thermal modeling with a partially illuminated primary body are found in \cite{furukawa_practical_1992}. Implications of the low altitude and uniform illumination assumptions are further discussed in Section \ref{sec: SATMO Validation} of this work.

Another set of factors is defined to scale the solar and albedo heating intensity based on the spacecraft's geometry and angle from the subsolar point of the orbit. The scaling factors are estimated in Eq.~\eqref{eqn: F_sol} and Eq.~\eqref{eqn: F_alb} based on \cite{rickman_introduction_2014}.

\begin{equation}
\label{eqn: F_sol}
\mathcal{F}_{sol{\to}_i} =
\begin{cases}
    \cos(\xi), & i = zen \text{ and } (\theta \leq 90^\circ \text{ or }   \theta \geq 270^\circ) \\
    -\cos(\xi), & i = nad \text{ and } (90^\circ \leq \theta \leq \theta_1 \text{ or } \theta_2 \leq \theta \leq 270^\circ) \\
    -\sin(\theta)\cos(\beta), & i = +v \text{ and } \theta_2 \leq \theta \leq 360^\circ \\
    \sin(\theta)\cos(\beta), & i = -v \text{ and } 0^\circ \leq \theta \leq \theta_1 \\
    \lvert \sin(\beta) \rvert, & i = N \text{ and } \beta > 0^\circ \text{ and } (0^\circ \leq \theta \leq \theta_1 \text{ or } \theta_2 \leq \theta \leq 360^\circ) \\
    \lvert \sin(\beta) \rvert, & i = S \text{ and } \beta < 0^\circ \text{ and } (0^\circ \leq \theta \leq \theta_1 \text{ or } \theta_2 \leq \theta \leq 360^\circ) \\
    0, & \text{otherwise}
\end{cases}
\end{equation}

\begin{equation}
\label{eqn: F_alb}
\mathcal{F}_{alb{\to}_i} =
\begin{cases}
\cos(\xi), & i = nad,+v,-v,N,S \text{ and } (\xi \leq 90^\circ \text{ or } \xi \geq 270^\circ) \\
0, & \text{otherwise}
\end{cases}
\end{equation}
where the $\cos(\xi)$ terms in Eq.~\eqref{eqn: F_sol} and Eq.~\eqref{eqn: F_alb} capture the trailing-off effect of the solar intensity as the spacecraft moves away from the subsolar point of its orbit \cite{gilmore_spacecraft_2002}.

\section{SATMO Validation}
\label{sec: SATMO Validation}
This section of the work validates satellite absorbed heat fluxes and surface temperatures calculated by SATMO against those calculated by TD. The 1U CubeSat for SATMO validation is modeled as an orbiting box made from aluminum 6061-T6 with identical black body surfaces. Each surface has properties found in Table \ref{tab: SATMO validation satellite props}. SATMO supports unique user inputs for each surface like asymmetric internal heat loading, which is typical for spacecraft. However, this simple example assumes that 3 W of heat is generated by the internal electronics, adding a 0.5 W load to each surface at all times. A bang-bang heat control algorithm is included for each 1 W surface heater to switch on when a respective surface temperature drops to $\leq 0^\circ$C and to switch off when the temperature rises to $\geq 10^\circ$C. Body-mounted solar panels are not included in this example, so the effective surface properties $\alpha$ and $\varepsilon$ used in the SATMO thermal calculations are equivalent to those of the satellite face in Table \ref{tab: SATMO validation satellite props}.

\begin{table}[H]
\caption{\label{tab: SATMO validation satellite props} Satellite property inputs for the SATMO validation cases. Properties are identical for each of the six satellite faces. The conductance coefficient matrix for the satellite surfaces is found in Table \ref{tab: contact conductance}.}
\centering
\begin{tabular}{lc}
\hline
SATMO field & Input \\
\hline
\multicolumn{2}{c}{\textbf{Satellite face}} \\
Mass, kg & 0.25 \\
Area, \si[per-mode=symbol]{\m\squared} & 0.010 \\
Specific heat, \si[per-mode=symbol]{\J\per\kg\per\kelvin} & 896.0 \\
Absorptivity & 1.0 \\
Emissivity & 1.0 \\
Initial temperature, $^\circ$C & 20.0 \\
Internal heat load, W & 0.50 \\
\multicolumn{2}{c}{\textbf{Heater and heater control}} \\
Heater power, W & 1.0 \\
Bang-bang control & On \\
On trigger, $^\circ$C & 0.0 \\
Off trigger, $^\circ$C & 10.0 \\
\multicolumn{2}{c}{\textbf{Solar panel}} \\
Mounting & None \\
Face coverage, \% & N/A \\
Efficiency & N/A \\
Absorptivity & N/A \\
Emissivity & N/A \\
\hline
\end{tabular}
\end{table}

The conductance coefficient matrix used in SATMO validation is presented in Table \ref{tab: contact conductance}, where a uniform value $K = 0.12$ \si[per-mode=symbol]{\watt\per\kelvin} is applied between all contacting surfaces. The matrix is symmetric, and its block diagonal entries are zero, indicating no direct conduction between opposing surface pairs. The absolute conduction coefficient value of 0.12 $\si[per-mode=symbol]{\watt\per\kelvin}$ comes from a conservative estimate for aluminum 6061-T6, following \cite{versteeg_preliminary_2018}, assuming a contact area of \num{4.0e-4} \si[per-mode=symbol]{\m\squared} between adjacent surfaces.

\begin{table}[H]
\caption{\label{tab: contact conductance} Conductance matrix between surfaces of the satellite in this work, in $\text{{W/K}}$.}
\centering
\begin{tabular}{lcccccc}
\hline
 & Zenith & Nadir & Forward & Aft & North & South \\\hline
Zenith  & 0 & 0 & 0.12 & 0.12 & 0.12 & 0.12 \\
Nadir   & 0 & 0 & 0.12 & 0.12 & 0.12 & 0.12 \\
Forward & 0.12 & 0.12 & 0 & 0 & 0.12 & 0.12 \\
Aft     & 0.12 & 0.12 & 0 & 0 & 0.12 & 0.12 \\
North   & 0.12 & 0.12 & 0.12 & 0.12 & 0 & 0 \\
South   & 0.12 & 0.12 & 0.12 & 0.12 & 0 & 0 \\
\hline
\end{tabular}
\end{table}

\begin{figure}[H]
\centering
\includegraphics[scale=0.9]{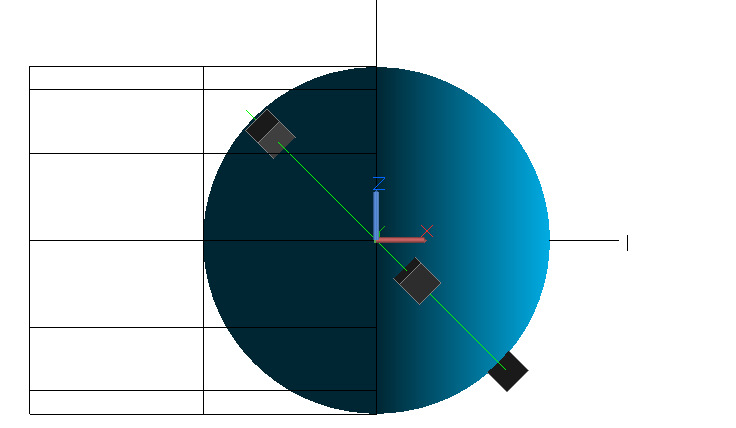}
\caption{Orbit plane side view of Test Case 2 from Table \ref{tab: test cases} modeled in TD with multiple satellite positions shown. The grid lines represent the shadow zone, and the size of the spacecraft is enlarged for better display.}
\label{fig: TD Screen Capture} 
\end{figure}

Table \ref{tab: test cases} presents the test matrix used to benchmark SATMO against TD. The test cases allow for result comparisons across the thermal models with the same input satellite characteristics, independently varying the beta angle, orbit altitude, and primary body. $\beta$ is varied for an Earth-orbiting satellite at $0^\circ$, $45^\circ$, and $90^\circ$, representing low-, medium-, and high-$\beta$ scenarios, respectively. $h$ is varied for an Earth-orbiting satellite at 400 km, 800 km, and 35,786 km, representing Low Earth Orbit (LEO), Sun-synchronous orbit (SSO), and geosynchronous (GEO) orbit altitudes, respectively. While the SATMO program functions the same regardless of the primary body input, running simulations with various planets allows the model to be tested with the different environmental heating parameters presented in Table \ref{tab: planet data}. Environmental heating rate constants for other bodies under various orbital scenarios can be found in \cite{gilmore_spacecraft_2002}. Radius, mass, equatorial inclination, and J2 values from Table \ref{tab: planet data} are taken from \cite{williams_planetary_2025} and environmental heating parameters are from \cite{gilmore_spacecraft_2002}. Figure \ref{fig: TD Screen Capture} displays Test Case 2 as modeled in TD, with multiple spacecraft positions displayed along the orbit.

\begin{table}[H]
\caption{\label{tab: test cases} Test cases used for SATMO validation. All simulations are performed using time step \boldmath$\Delta t = 1.0$ s}
\centering
\begin{tabular}{lcccc}
\hline
    Test Case & Primary body & Beta angle, deg & Orbit altitude, km & Simulation duration, s \\\hline
1  & Earth & 0 & 400 & 27,768.1\\
2  & Earth & 45 & 400 & 27,768.1 \\
3  & Earth & 45 & 800 & 30,262.0\\
4  & Earth & 45 & 35,786 & 430,819 \\
5  & Earth & 90 & 400 & 27,768.1 \\
6  & Mars & 45 & 400 & 35,506.5\\
7  & Venus & 45 & 400 & 28,564.3\\
\hline
\end{tabular} \end{table}

\begin{table}[H]
\caption{\label{tab: planet data} Planetary data constants used in SATMO validation, with values adopted from \cite{williams_planetary_2025} and \cite{gilmore_spacecraft_2002}.}
\centering
\begin{tabular}{lccc}
\hline
    & Venus & Earth & Mars \\\hline
Radius, km  & 6,051.800 & 6,378.137 & 3,396.200 \\
Mass, kg  & \num{4.8673e+24} & \num{5.9722e+24} & \num{6.4169e+23} \\
Equatorial inclination, deg  & 2.64 & 23.44 & 25.19 \\
J2 constant  & \num{4.45800e-6}   & \num{1.08263e-3} & \num{1.96045e-3}\\
Solar flux, \si[per-mode=symbol]{\watt\per\m\squared}  & 2,759 & 1,414 & 717 \\
Albedo factor  & 0.82 & 0.40 & 0.29 \\
IR flux, \si[per-mode=symbol]{\watt\per\m\squared} & 153 & 218 & 315 \\
\hline
\end{tabular} \end{table}

Absorbed heat flux comparisons are performed over one complete orbit, while surface temperatures are compared over five orbits, allowing a quasi-steady-state temperature profile to develop. All SATMO validation simulations are performed using $\Delta t = 1.0$ s. The heating rates and temperature comparison results are summarized in Table \ref{tab: flux and temp differences}. Notably, Table \ref{tab: flux and temp differences} reveals that the approaches taken to calculate the solar flux across SATMO and TD yield identical results, and the IR heating rates are very similar between models. Heating rates due to albedo (i.e., "albedo flux") are the least consistent across the two models, but SATMO is considered validated for the provided test cases with the best case, maximum, absolute temperature discrepancy of $0.02^\circ$C between models and the worst case, maximum, absolute temperature discrepancy of $1.17^\circ$C.

\begin{table}[H]
\caption{\label{tab: flux and temp differences} Maximum difference between SATMO and TD calculations for absorbed surface environmental heating rates and temperatures based on satellite properties in Table \ref{tab: SATMO validation satellite props} and planetary data in Table \ref{tab: planet data}. Test case descriptions are found in Table \ref{tab: test cases}. Solar flux calculations across the models are identical.}
\centering
\begin{tabular}{lcccccc}
\hline
     & Zenith & Nadir & Forward & Aft & North & South \\\hline
    \multicolumn{7}{c}{\textbf{Test Case 1 (400-km-altitude Earth Orbit, \boldmath$\beta = 0^\circ$)}}\\
Solar flux, \si[per-mode=symbol]{\watt\per\m\squared} & 0.00 & 0.00 & 0.00 & 0.00 & 0.00 & 0.00\\
Albedo flux, \si[per-mode=symbol]{\watt\per\m\squared} & 0.00 & 11.43 & 13.56 & 13.45 & 4.26 & 4.30 \\
IR flux, \si[per-mode=symbol]{\watt\per\m\squared} & 0.00 & 0.01 & -0.04 & -0.05 & -0.05 & -0.45 \\
Temperature, $^\circ$C & -0.05 & -0.07 & -0.24 & -0.18 & -0.06 & -0.06 \\

    \multicolumn{7}{c}{\textbf{Test Case 2 (400-km-altitude Earth Orbit, \boldmath$\beta = 45^\circ$)}}\\
    Solar flux, \si[per-mode=symbol]{\watt\per\m\squared} & 0.00 & 0.00 & 0.00 & 0.00 & 0.00 & 0.00\\
        Albedo flux, \si[per-mode=symbol]{\watt\per\m\squared} & 0.00 & 11.53 & 10.17 & 10.06 & 10.18 & -11.52 \\
IR flux, \si[per-mode=symbol]{\watt\per\m\squared} & 0.00 & 0.06 & 0.26 & -0.04 & 0.00 & 0.52 \\
Temperature, $^\circ$C & 0.04 & 0.06 & 0.16 & 0.13 & 0.20 & 0.19 \\

    \multicolumn{7}{c}{\textbf{Test Case 3 (800-km-altitude Earth Orbit,  \boldmath$\beta = 45^\circ$)}}\\
Solar flux, \si[per-mode=symbol]{\watt\per\m\squared} & 0.00 & 0.00 & 0.00 & 0.00 & 0.00 & 0.00\\
Albedo flux, \si[per-mode=symbol]{\watt\per\m\squared} & 0.00 & 17.85 & 12.3 & 12.43 & 12.31 & -13.83 \\
IR flux, \si[per-mode=symbol]{\watt\per\m\squared} & 0.00 & 0.03 & 0.43 & -0.64 & 0.11 & 0.16 \\
Temperature, $^\circ$C & 0.06 & 0.13 & -0.22 & 0.13 & 0.23 & -0.28 \\

    \multicolumn{7}{c}{\textbf{Test Case 4 (35,786-km-altitude Earth Orbit,  \boldmath$\beta = 45^\circ$)}}\\
Solar flux, \si[per-mode=symbol]{\watt\per\m\squared} & 0.00 & 0.00 & 0.00 & 0.00 & 0.00 & 0.00\\
Albedo flux, \si[per-mode=symbol]{\watt\per\m\squared} & 0.00 & 2.49 & 0.17 & 0.17 & 0.17 & -0.23 \\
IR flux, \si[per-mode=symbol]{\watt\per\m\squared} & 0.00 & 0.00 & 0.00 & 0.00 & 0.00 & 0.00 \\
Temperature, $^\circ$C & -0.06 & -0.01 & -0.07 & 0.06 & 0.06 & -0.07 \\

    \multicolumn{7}{c}{\textbf{Test Case 5 (400-km-altitude Earth Orbit,  \boldmath$\beta = 90^\circ$)}}\\
Solar flux, \si[per-mode=symbol]{\watt\per\m\squared} & 0.00 & 0.00 & 0.00 & 0.00 & 0.00 & 0.00\\
Albedo flux, \si[per-mode=symbol]{\watt\per\m\squared}     & 0.00 & 11.62 & 4.41 & 4.38 & 13.70 & 0.00 \\
IR flux, \si[per-mode=symbol]{\watt\per\m\squared} & 0.00 & -0.07 & -0.38 & -0.16 & 0.60 & -0.18 \\
Temperature, $^\circ$C & 0.96 & 1.17 & 1.05 & 1.04 & 1.17 & 0.96 \\

    \multicolumn{7}{c}{\textbf{Test Case 6 (400-km-altitude Mars Orbit,  \boldmath$\beta = 45^\circ$)}}\\
Solar flux, \si[per-mode=symbol]{\watt\per\m\squared} & 0.00 & 0.00 & 0.00 & 0.00 & 0.00 & 0.00\\
Albedo flux, \si[per-mode=symbol]{\watt\per\m\squared} & 0.00 & 6.34 & 4.52 & 4.52 & 4.54 & -5.06 \\
IR flux, \si[per-mode=symbol]{\watt\per\m\squared} & 0.00 & -0.12 & 0.03 & 0.29 & 0.72 & -0.56 \\
Temperature, $^\circ$C & -0.02 & 0.05 & -0.08 & 0.07 & 0.08 & -0.11 \\

    \multicolumn{7}{c}{\textbf{Test Case 7 (400-km-altitude Venus Orbit,  \boldmath$\beta = 45^\circ$)}}\\
Solar flux, \si[per-mode=symbol]{\watt\per\m\squared} & 0.00 & 0.00 & 0.00 & 0.00 & 0.00 & 0.00\\
Albedo flux, \si[per-mode=symbol]{\watt\per\m\squared}         & 0.00 & 47.62 & 41.08 & 41.48 & 41.61 & -46.60 \\
IR flux, \si[per-mode=symbol]{\watt\per\m\squared} & 0.00 & -0.17 & -0.02 & 0.22 & 0.18 & 0.00 \\
Temperature, $^\circ$C & -0.12 & 0.22 & -0.66 & 0.47 & 0.67 & -0.82 \\
\hline
\end{tabular} \end{table}

\begin{figure}[H]
\centering
\includegraphics[width=5.85in, height=3.897in]{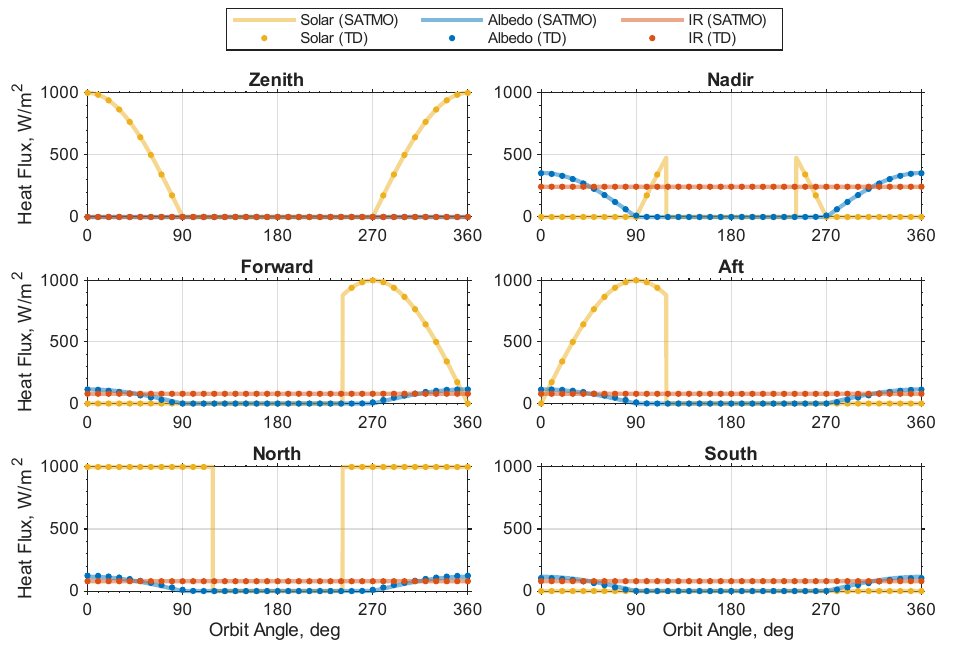}
\caption{Absorbed environmental heat fluxes on surfaces calculated by SATMO and TD for Test Case 2 based on satellite properties in Table \ref{tab: SATMO validation satellite props} and planetary data in Table \ref{tab: planet data}. Test case descriptions are found in Table \ref{tab: test cases}.}
\label{fig: Raw Fluxes}
\end{figure}

\begin{figure}[H]
\centering
\includegraphics[width=5.85in, height=3.897in]{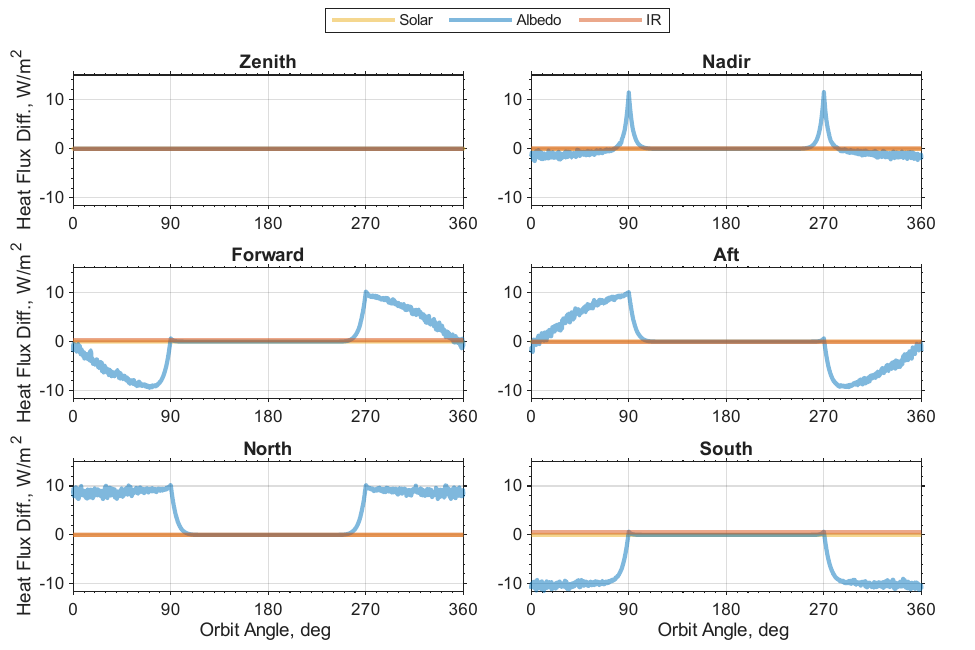}
\caption{Differences in calculated absorbed environmental heat fluxes on surfaces between SATMO and TD for Test Case 2 based on satellite properties in Table \ref{tab: SATMO validation satellite props} and planetary data in Table \ref{tab: planet data}. Test case descriptions are found in Table \ref{tab: test cases}.}
\label{fig: Flux Differences} 
\end{figure}

\begin{figure}[H]
\centering
\includegraphics[width=5.85in, height=3.897in]{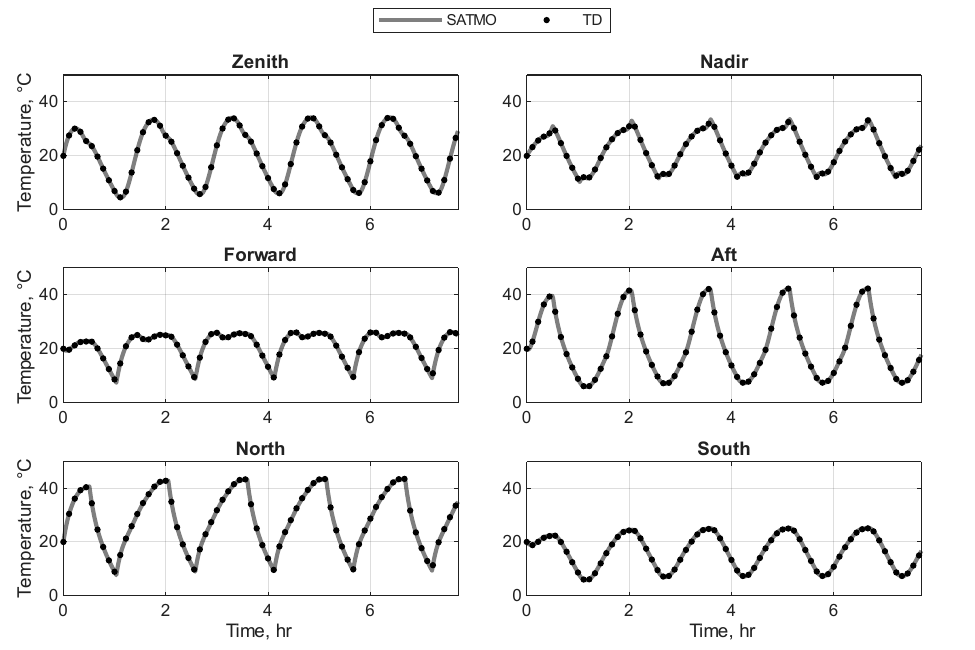}
\caption{Surface temperatures calculated by SATMO and TD for Test Case 2 based on satellite properties in Table \ref{tab: SATMO validation satellite props} and planetary data in Table \ref{tab: planet data}. Test case descriptions are found in Table \ref{tab: test cases}.}
\label{fig: Temp Comparison} 
\end{figure}

\begin{figure}[H]
\centering
\includegraphics[width=5.85in, height=3.897in]{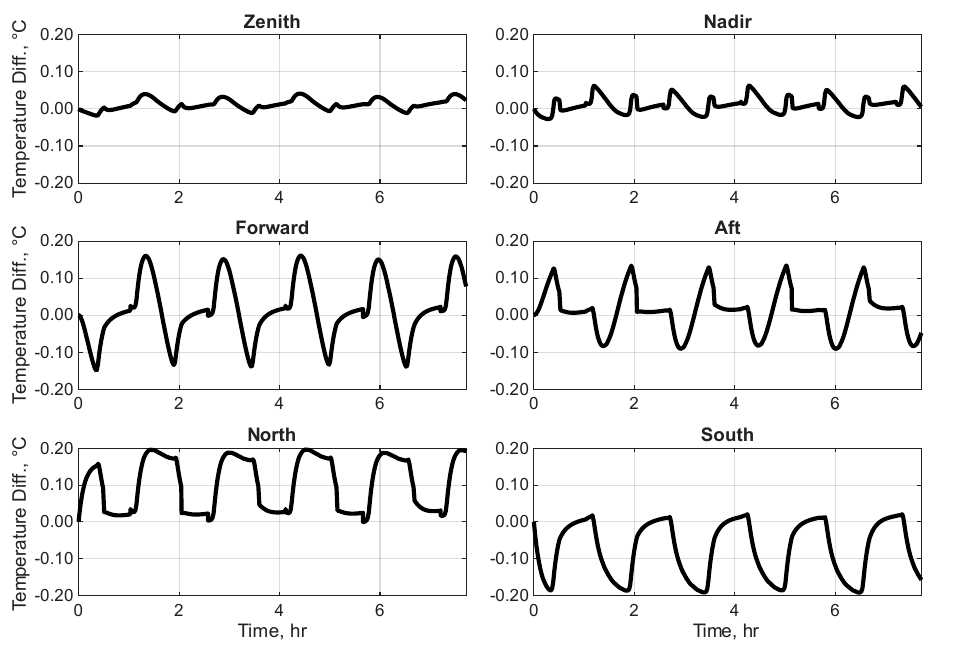}
\caption{Differences in calculated surface temperatures between SATMO and TD for Test Case 2 based on satellite properties in Table \ref{tab: SATMO validation satellite props} and planetary data in Table \ref{tab: planet data}. Test case descriptions are found in Table \ref{tab: test cases}.}
\label{fig: Temp Difference} 
\end{figure}

Sample outputs from Test Case 2 are provided in Figs.~\ref{fig: Raw Fluxes}--\ref{fig: Temp Difference} to help visualize the behavior of the models. As observed in Fig.~\ref{fig: Raw Fluxes}, the heating rates calculated by SATMO follow similar trends to those calculated by TD. However, Fig.~\ref{fig: Flux Differences} reveals noticeable discrepancies in albedo heating rates. For the Nadir surface, the albedo flux difference spikes near the terminator regions, around $90^\circ$ and $270^\circ$. This is because SATMO approximates there to be no albedo heating contribution at the terminator per Eq.~\eqref{eqn: F_alb}. In reality, albedo heating rates do not vanish right at the terminator \cite{vanoutryve_thermal_2008}, but rather they trail off through more complex albedo-shadow umbra/penumbra modeling. The near-terminator albedo behavior is exacerbated for Test Case 5 where $\beta = 90^\circ$ because SATMO assumes no albedo heating contributions for the Nadir, Forward, Aft, and North surface, while TD models a non-zero constant albedo flux for these surfaces over an entire orbit. Test Case 5 has the largest temperature discrepancy between any surface modeled by SATMO and TD, with TD calculating North and Nadir surface temperatures $1.17^\circ$C greater than SATMO.

In terms of the Forward-Aft and the North-South surface pairs, differences in albedo heating rates appear to be almost equal and opposite to one another. In SATMO, these four surfaces are modeled to receive the same amount of albedo flux due to SATMO's assumption of a uniformly illuminated primary body beneath each satellite surface. The discrepancies in albedo heating rates arise because TD accounts for the non-uniform illumination of the primary body as viewed from each satellite surface. For example, at the subsolar point---located south of the equator for $\beta = 45^\circ$ in Fig.~\ref{fig: Flux Differences}---the North surface in TD views a fully illuminated portion of Earth, while the South surface views a region that is partially illuminated and partially shadowed. These discrepancies are exacerbated in Test Case 7 due to Venus' high albedo factor ($a_p = 0.82$). The largest absolute difference in heating rates away from the terminator region is $46.60$ \si[per-mode=symbol]{\watt\per\m\squared} on the South surface, which results in less than 0.5 W of additional heat transfer onto the South surface of a 1U CubeSat.

An interesting trend emerges when comparing the albedo flux discrepancies between SATMO and TD across Test Cases 2, 3, and 4 in Table \ref{tab: flux and temp differences}, in which only orbit altitude varies. Test Case 3 ($h = 800$ km) exhibits larger albedo flux discrepancies than Test Case 2 ($h = 400$ km), while Test Case 4 ($h = 35,786)$ presents a much smaller discrepancy than Test Case 2 ($h = 400$ km). This suggests that for a given orbital configuration, there exists some altitude $h$ at which the local view factor to the primary body becomes sufficiently small enough that it minimizes the albedo heating effects caused by the uniform illumination assumption. Meanwhile, at high altitudes, the cylindrical shadow approximation begins to break down \cite{rickman_simplified_2002}, making it difficult to precisely define the maximum orbit altitude for which SATMO's modeling approach remains valid. As shown in Fig. \ref{fig: Flux Differences}, the albedo flux calculated by TD between Forward-Aft and North-South surface pairs tend to average out to those calculated by SATMO, contributing to the overall agreement in temperature profiles across the test cases since the surfaces are thermally connected via conduction. The general agreement between temperatures calculated by SATMO and TD is illustrated in Fig. \ref{fig: Temp Comparison} for Test Case 2, while the corresponding temperature discrepancies are shown in Fig. \ref{fig: Temp Difference} and are attributed to differences in environmental heat fluxes.

\begin{figure}[H]
\centering
\includegraphics[width=5.525in, height=3.1875in]{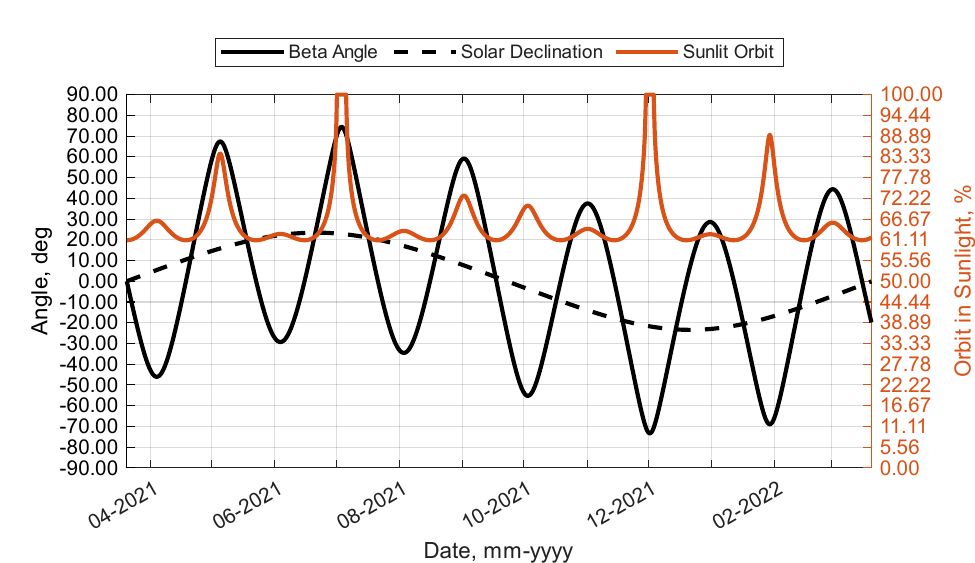}
\caption{Beta angle evolution for an example ISS orbit with \boldmath$h = 400$ km, \boldmath$i_{sat} = 51.6^\circ$, and initial \boldmath$\Omega = 0^\circ$ beginning on March 20th, 2021 at 09:37 UTC and lasting one calendar year. ISS orbital data are from \cite{kuzlu_modeling_2021} and \cite{boushon_thermal_2018}.} 
\label{fig: Earth ISS Beta Evolution Validation} 
\end{figure}

\begin{figure}[H]
\centering
\includegraphics[width=5.525in, height=3.1875in]{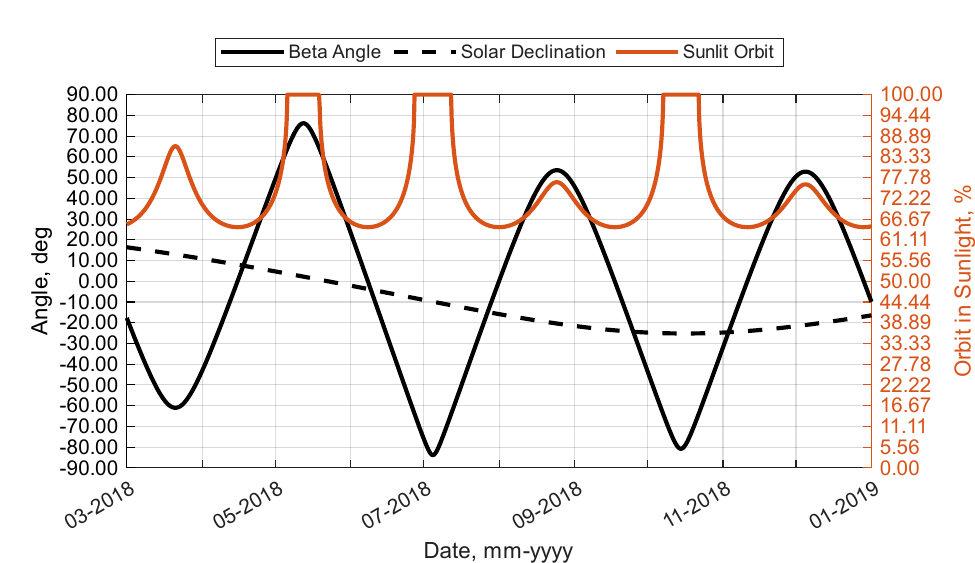}
\caption{Beta angle evolution for an example TGO orbit with \boldmath$h = 385$ km, \boldmath$i_{sat} = 74^\circ$, and initial \boldmath$\Omega = 115^\circ$ beginning on March 1st, 2018 at 0:00 UTC and lasting 10 calendar months. TGO orbital data come from \cite{geiger_long_2018}.}
\label{fig: MARS TGO Beta Evolution Validation} 
\end{figure}

An important feature of SATMO's specific analysis mode is its ability to present $\beta$ evolution over calendar date to help identify when and for how long extreme $\beta$ values are expected to occur. Figure \ref{fig: Earth ISS Beta Evolution Validation} and Fig.~\ref{fig: MARS TGO Beta Evolution Validation} are provided as visual references to verify that the approach in Eqs.~\eqref{eqn: betaCrit}--\eqref{eqn: solarDeclination} performs as expected. Figure \ref{fig: Earth ISS Beta Evolution Validation} is generated with an example International Space Station (ISS) orbit \cite{kuzlu_modeling_2021, boushon_thermal_2018} with $h = 400$ km, $i_{sat} = 51.6^\circ$, and initial $\Omega = 0^\circ$ beginning on March 20th, 2021 at 09:37 UTC---Earth's vernal equinox---and lasting one calendar year. Figure \ref{fig: MARS TGO Beta Evolution Validation} is generated with an example ExoMars Trace Gas Orbiter (TGO) orbit \cite{geiger_long_2018} with $h = 385$ km, $i_{sat} = 74^\circ$, and initial $\Omega = 115^\circ$ beginning on March 1st, 2018 at 0:00 UTC and lasting 10 calendar months. These plots can be cross-referenced with the $\beta$ evolution plots found in \cite{kuzlu_modeling_2021} and \cite{boushon_thermal_2018} for the ISS example orbit and \cite{geiger_long_2018} for the TGO example orbit. Minor discrepancies in results can be attributed to alternative methods for calculating $\beta$ over time, but the results are considered qualitatively consistent with an acceptable order of accuracy.

\section{Example use of SATMO}
This section presents an example use case of SATMO and its outputs available to users. Say, for instance, a mission design team wants to use SATMO to estimate the thermal behavior of a 1U CubeSat orbiting Mars with simulation parameters in Table \ref{tab: SATMO example simulation props} and identical surface properties in Table \ref{tab: SATMO example satellite props}. The satellite in this section has the same properties as the arbitrary 1U CubeSat from Section \ref{sec: SATMO Validation} of this work, with the exceptions of: (1) the absence of any heaters and (2) the presence of body-mounted solar panels on each surface. The presence of solar panels modifies the effective surface properties used in SATMO's thermal calculations. For the hot case, $\alpha = 0.892$ and $\varepsilon = 0.820$; for the cold case, $\alpha = 0.622$ and $\varepsilon = 0.820$. The satellite also has the same total internal heat generation and conductance coefficient matrix (Table \ref{tab: contact conductance}) as the CubeSat described in Section \ref{sec: SATMO Validation}.

\begin{table}[H]
\caption{\label{tab: SATMO example simulation props} Simulation parameters of a Mars-orbiting CubeSat for demonstrating an example SATMO use case. Mars environmental heating values are adopted from \cite{gilmore_spacecraft_2002}.}
\centering
\begin{tabular}{lc}
\hline
\textbf{SATMO field} & \textbf{Input} \\
\hline
\multicolumn{2}{c}{\textbf{Analysis-independent}} \\
Simulation duration, s & 28,237 \\
Time step, s & 10 \\
Primary body & Mars \\
Orbit altitude, km & 385 \\[3pt]

\multicolumn{2}{c}{\textbf{Specific analysis}} \\
Epoch, UTC & 17-Aug-2028 00:00:00 \\
Orbital inclination, deg & 74 \\
RAAN, deg & 0 \\
Beta angle evolution, calendar days & 687\\
Beta angle (hot case), deg & 63.92 \\
Beta angle (cold case), deg & 0 \\[3pt]

\multicolumn{2}{c}{\textbf{Generic analysis}} \\
Lower beta angle bound, deg & -90 \\
Beta angle increment, deg & 5 \\
Upper beta angle bound, deg & 90 \\[3pt]

\multicolumn{2}{c}{\textbf{Primary body}} \\
Radius, km & 3,396.20 \\
Mass, kg & \num{6.4169e+23} \\
Equatorial inclination, deg & 25.19 \\
J2 constant & \num{1.96045e-3} \\[3pt]

\multicolumn{2}{c}{\textbf{Environmental heating}} \\

 & Hot case \quad Cold case \\
Solar flux, \si[per-mode=symbol]{\watt\per\meter\squared} & 717 \quad 493 \\
Albedo factor & 0.29 \quad 0.29 \\
IR Planetshine, \si[per-mode=symbol]{\watt\per\meter\squared} & 470 \quad 315 \\
\hline
\end{tabular}
\end{table}

\begin{table}[H]
\caption{\label{tab: SATMO example satellite props} Properties of a Mars-orbiting CubeSat for demonstrating an example SATMO use case. Properties are identical for each of the six satellite faces. The conductance coefficient matrix for the satellite surfaces is found in Table \ref{tab: contact conductance}.}
\centering
\begin{tabular}{lc}
\hline
SATMO field & Input \\
\hline
\multicolumn{2}{c}{\textbf{Satellite face}} \\
Mass, kg & 0.25 \\
Area, \si[per-mode=symbol]{\m\squared} & 0.010 \\
Specific heat, \si[per-mode=symbol]{\J\per\kg\per\kelvin} & 896.0 \\
Absorptivity & 1.0 \\
Emissivity & 1.0 \\
Initial temperature, $^\circ$C & 20.0 \\
Internal heat load, W & 0.50 \\
\multicolumn{2}{c}{\textbf{Heater and heater control}} \\
Heater power, W & 0.0 \\
Bang-bang control & Off \\
On trigger, $^\circ$C & N/A \\
Off trigger, $^\circ$C & N/A \\
\multicolumn{2}{c}{\textbf{Solar panel}} \\
Mounting & Body-mounted \\
Face coverage, \% & 90.0 \\
Efficiency & 0.30 \\
Absorptivity & 0.88 \\
Emissivity & 0.80 \\
\hline
\end{tabular}
\end{table}

Initially, the team runs SATMO in the generic analysis mode for four full orbits with $\Delta t = 10$ s across the full range of discrete $\beta$ scenarios using increments of $5^\circ$ to visualize the full thermal ranges that their spacecraft could experience. Table \ref{tab: SATMO example simulation props} includes the hot and cold case parameters for calculating the environmental heating rates for this Mars satellite.

\begin{figure}[H]
\centering
\includegraphics{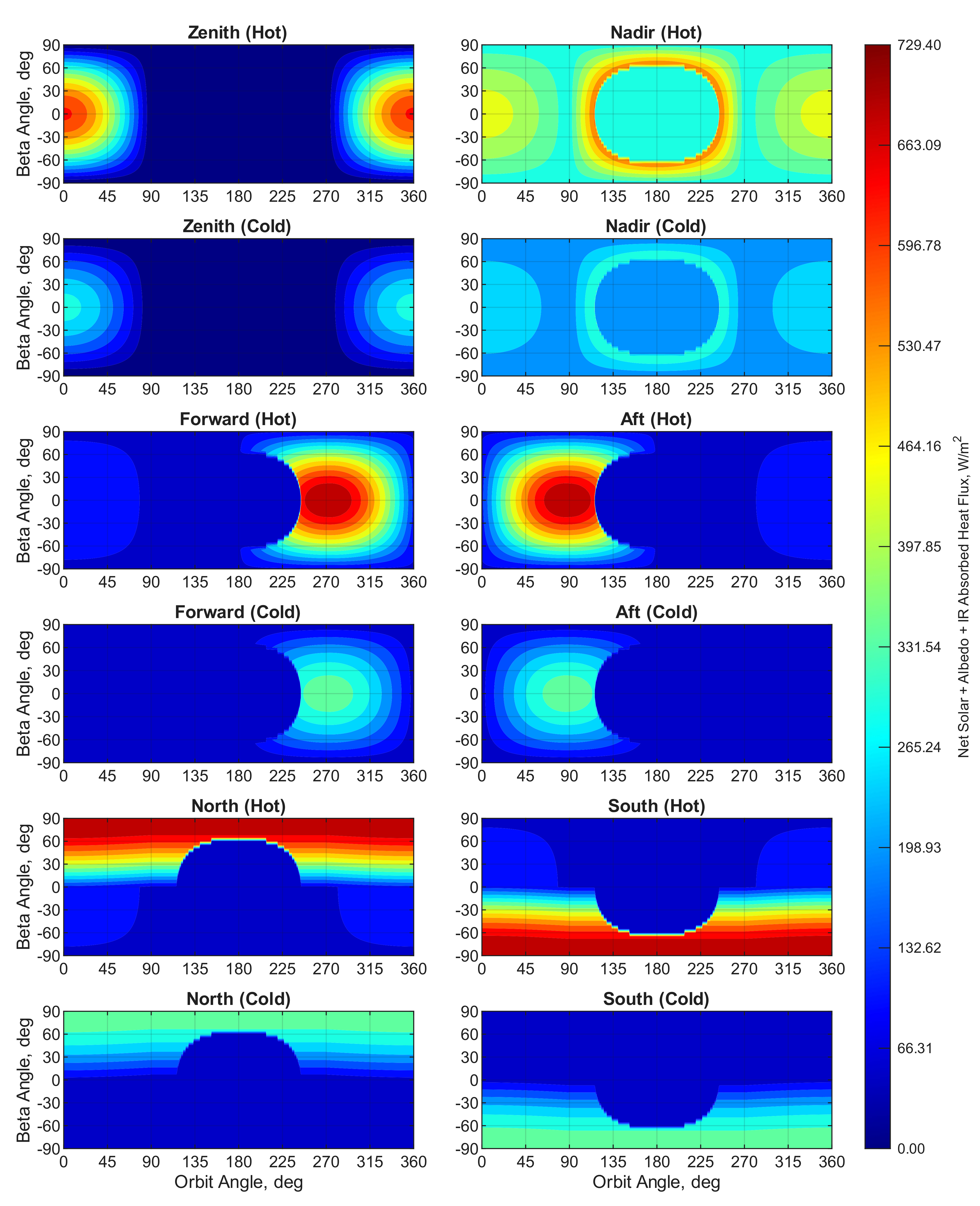}

\caption{Net total environmental heat fluxes absorbed by an example CubeSat orbiting Mars in SATMO's generic analysis mode (across a range of beta angles) over one orbital period (7,059.25 s). Reduction in net total absorbed heat fluxes occurs for a given surface as it is oriented away from the heat sources and/or as the satellite enters Mars' shadow zone. Simulations parameters are provided in Table \ref{tab: SATMO example simulation props} and Table \ref{tab: SATMO example satellite props}.}
\label{fig: Mars Example General Absorbed Fluxes} 
\end{figure}

The thermal output produced in Fig.~\ref{fig: Mars Example General Absorbed Fluxes} maps out the net environmental heat fluxes absorbed by each satellite surface from solar, albedo, and IR planetshine contributions. Note that the use of `net' here refers to the heat flux that is absorbed and contributes to temperature changes, not electrical power generation. Solar heating increases for the Zenith, Nadir, Forward, and Aft surfaces as $\beta$ approaches zero, driving up the absorbed heat flux when the CubeSat is outside Mars' shadow zone, as shown in Fig.~\ref{fig: Mars Example General Absorbed Fluxes}. Conversely, the North surface receives greater solar heating as $\beta$ deviates from zero, with the South surface exhibiting mirrored behavior. The Forward and Aft surfaces' absorbed heat fluxes also mirror each other due to their opposing locations on the CubeSat. SATMO users can view the raw incident heat fluxes on their satellite surfaces as a function of orbit angle by temporarily removing any solar panels and setting each surface's $\alpha = \varepsilon = 1$.

\begin{figure}[H]
\centering
\includegraphics{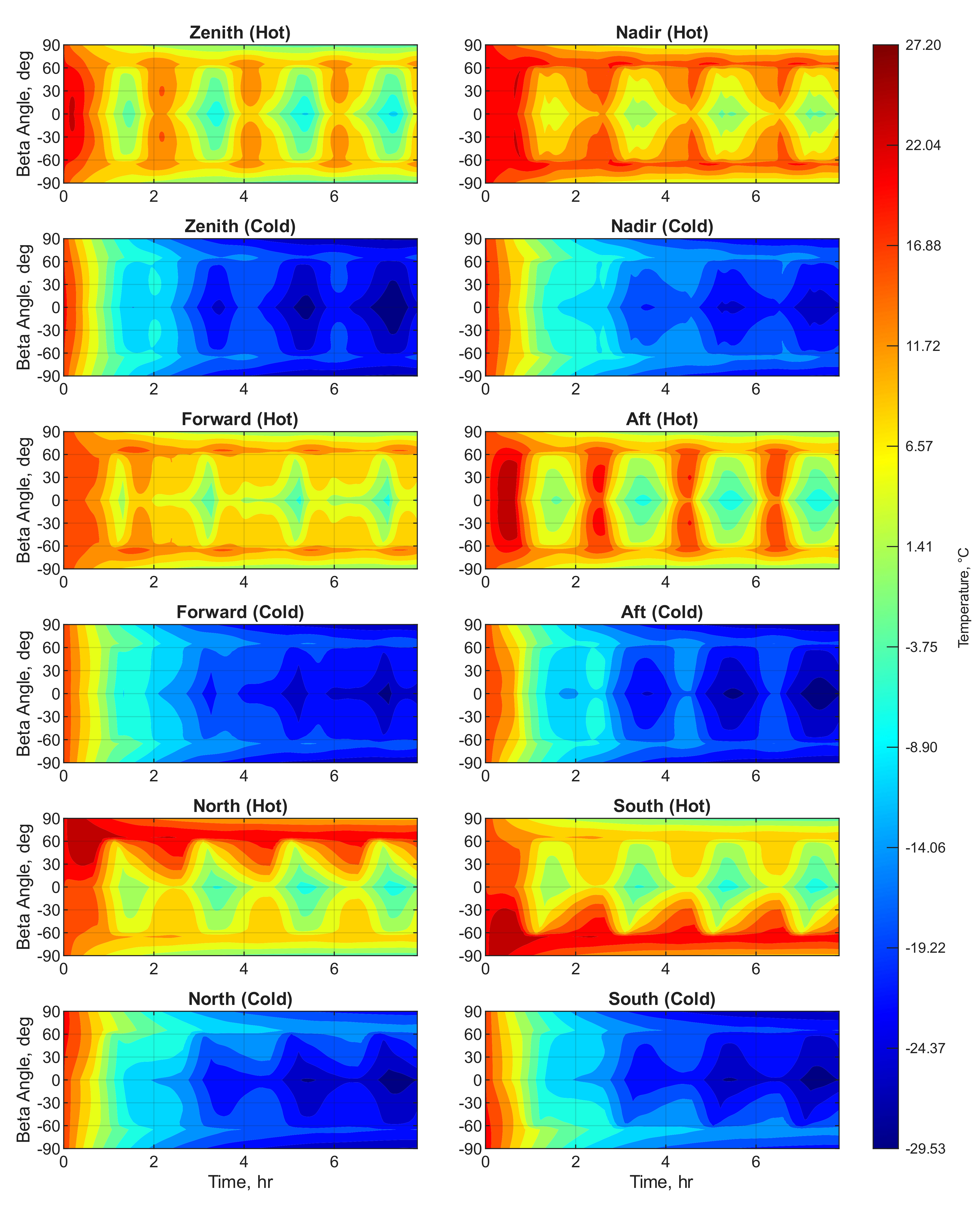}
\caption{Temperatures of an example CubeSat orbiting Mars in SATMO's generic analysis mode (across a range of beta angles) over four orbital periods (28,237 s). Simulations parameters are provided in Table \ref{tab: SATMO example simulation props} and Table \ref{tab: SATMO example satellite props}.}
\label{fig: Mars Example General Panel Temps} 
\end{figure}

\begin{figure}[H]
\centering
\includegraphics{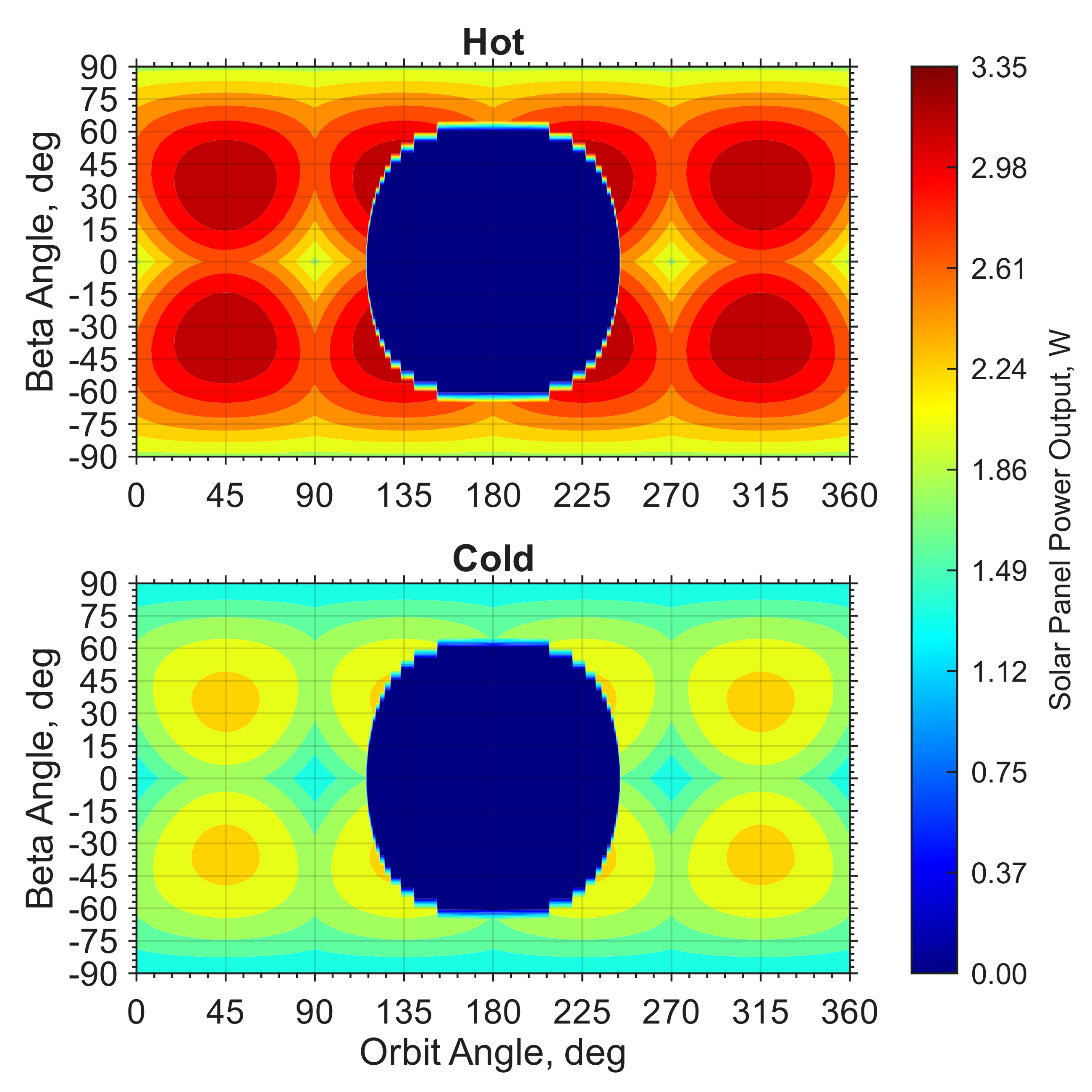}
\caption{Total instantaneous solar panel power output for an example CubeSat orbiting Mars in SATMO's generic analysis mode (across a range of beta angles) over one orbital period (7,059.25 s). Simulation parameters are provided in Table \ref{tab: SATMO example simulation props} and Table \ref{tab: SATMO example satellite props}.}
\label{fig: Mars Example General Solar Panel Power} 
\end{figure}

Fig.~\ref{fig: Mars Example General Panel Temps} maps out the temperatures of each satellite surface. The team can use Fig.~\ref{fig: Mars Example General Absorbed Fluxes} and Fig.~\ref{fig: Mars Example General Panel Temps} to identify $\beta$ regions in which the highest and lowest heating scenarios and temperatures result for each satellite surface, respectively. The total instantaneous electrical power output by the solar panels over a full orbit is also provided in Fig.~\ref{fig: Mars Example General Solar Panel Power}, for example. As expected, the environmental heating rates, temperatures, and solar panel power outputs are generally the lowest on the dark side of Mars and highest on the Sun side. SATMO also outputs a table of the minimum and maximum temperatures that each satellite surface will experience over each hot and cold case beta angle analysis. This feature helps the team easily identify the temperature extrema of their satellite during operation, though the table is not presented here for brevity.

\begin{figure}[H]
\centering
\includegraphics[width=5.525in, height=3.1875in]{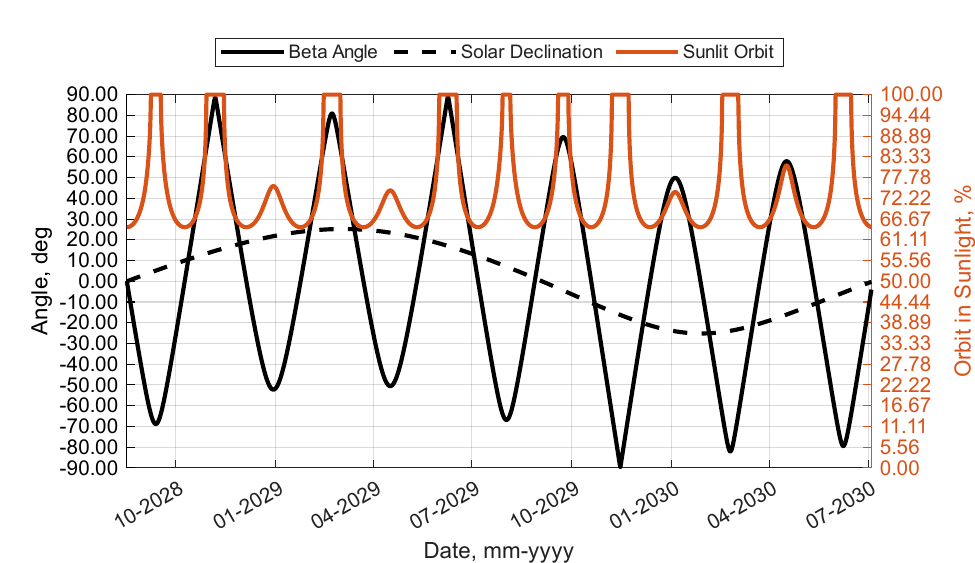}
\caption{Beta angle evolution and percentage of orbit in sunlight over 687 calendar days for an example CubeSat orbiting Mars. Simulation parameters are provided in Table \ref{tab: SATMO example simulation props} and Table \ref{tab: SATMO example satellite props}.}
\label{fig: Mars Example Beta Evolution} 
\end{figure}

\begin{figure}[H]
\centering
\includegraphics[width=5.525in, height=3.1875in]{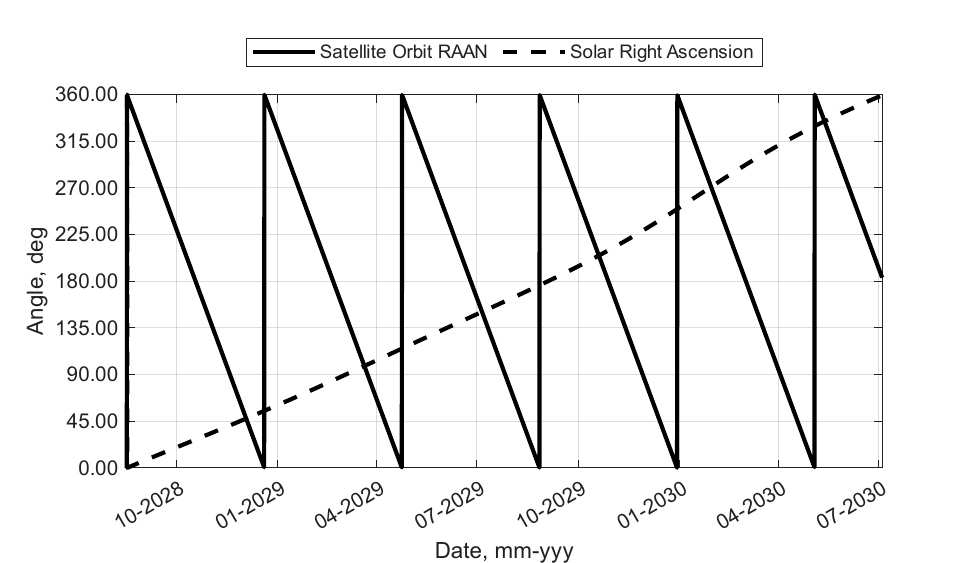}
\caption{RAAN and solar right ascension evolution over 687 calendar days for an example CubeSat orbiting Mars. Simulation parameters are provided in Table~\ref{tab: SATMO example simulation props} and Table~\ref{tab: SATMO example satellite props}.}
\label{fig: Mars Example RAAN and Solar RA Evolution} 
\end{figure}

Then, the team decides to run SATMO in specific analysis mode to hone in on the thermal management strategies for their spacecraft which is planned to enter its destination orbit---similar to the TGO \cite{geiger_long_2018}---with $h = 385$ km, $i_{sat} = 74^\circ$ and an initial $\Omega = 0^\circ$ on August 17, 2028 at 0:00 UTC, corresponding to an approximate future Martian vernal equinox. A hot-case $\beta =63.92^\circ$ is input into SATMO because that is the approximate beta angle at which the satellite becomes exposed to constant sunlight, per Eq.~\eqref{eqn: betaCrit} and Eq.~\eqref{eqn: f_E}. Conversely, a cold-case $\beta =0^\circ$ is input because that is when satellite orbit will spend the most time in eclipse. Upon running SATMO, Fig.~\ref{fig: Mars Example Beta Evolution} lets the team see when and for how long their spacecraft must reside in hot and cold cases over a full Martian year (687 calendar days). Figure \ref{fig: Mars Example Beta Evolution} reveals that the satellite orbit will experience constant sunlight for periods of up to 16 calendar days, with the first occurrence beginning on October 29th, 2028. A team may want to specify hot and cold case $\beta$ values based on a particular surface's exposure to sunlight, so SATMO also outputs each surface's percentage of orbit in sunlight as a function of calendar date, but this figure is not displayed in this work for brevity. Also, trends in $\Omega$ and $\Omega_s$ evolution are also made available to the user like in Fig.~\ref{fig: Mars Example RAAN and Solar RA Evolution}, which shows $\Omega$ returning to its initial value approximately every 94 calendar days.

\begin{figure}[H]
\centering
\includegraphics{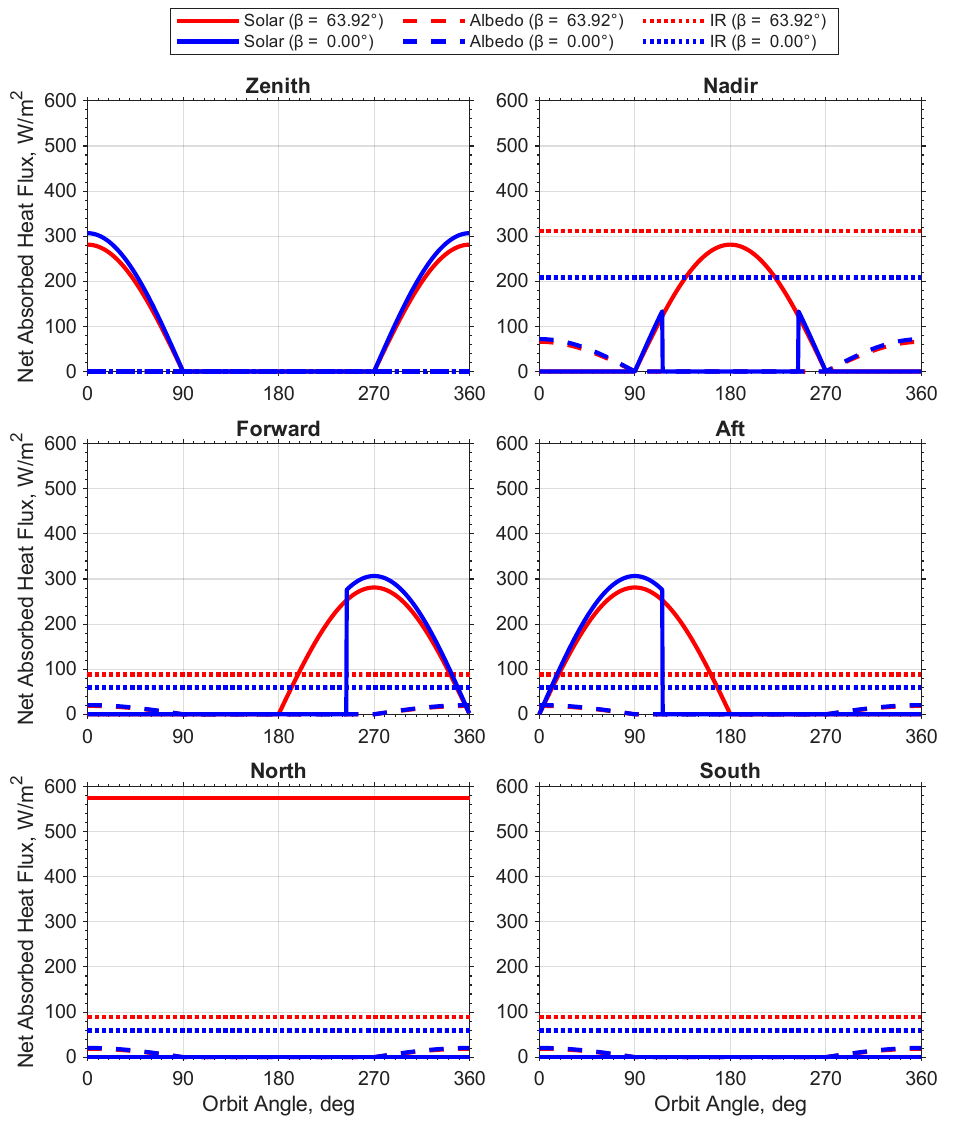}
\caption{Net individual environmental heat fluxes absorbed by an example CubeSat orbiting Mars output in SATMO's specific analysis mode (at distinct beta angles) over one orbital period (7,059.25 s). The environmental heat flux on each surface depends on its view factor to the primary body, its orientation relative to each heating source, and/or the satellite's angular position along its orbit. Simulation parameters are provided in Table~\ref{tab: SATMO example simulation props} and Table~\ref{tab: SATMO example satellite props}.}
\label{fig: Special Absorbed Fluxes (Individual)}
\end{figure}

\begin{figure}[H]
\centering
\includegraphics{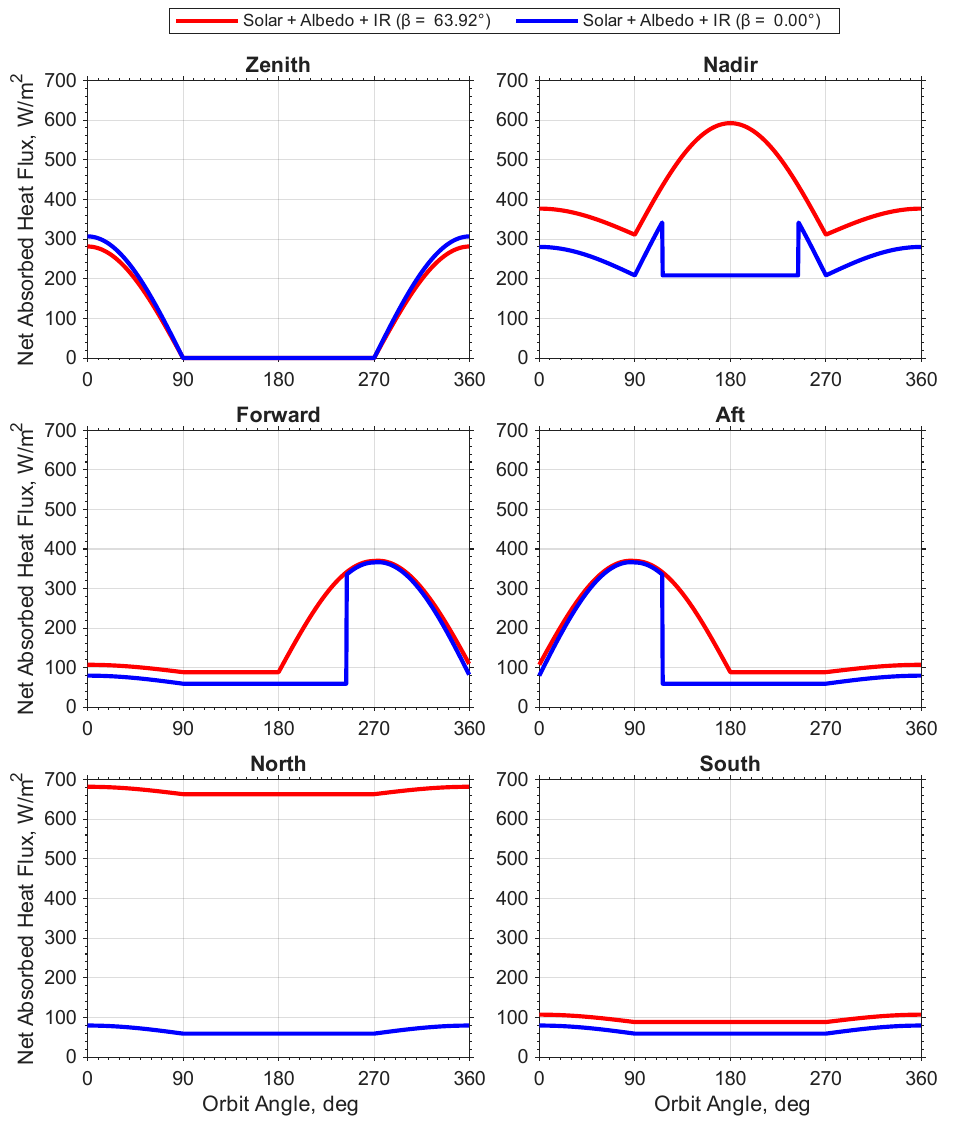}
\caption{Net total environmental heat fluxes absorbed by an example CubeSat orbiting Mars output in SATMO's specific analysis mode (at distinct beta angles) over one orbital period (7,059.25 s). In the hot case, the North surface absorbs the highest total heat flux over a full orbit, while in the cold case, the Nadir surface absorbs the most. Simulation parameters are provided in Table \ref{tab: SATMO example simulation props} and Table \ref{tab: SATMO example satellite props}.}
\label{fig: Special Absorbed Fluxes (Total)} 
\end{figure}

\begin{figure}[H]
\centering
\includegraphics{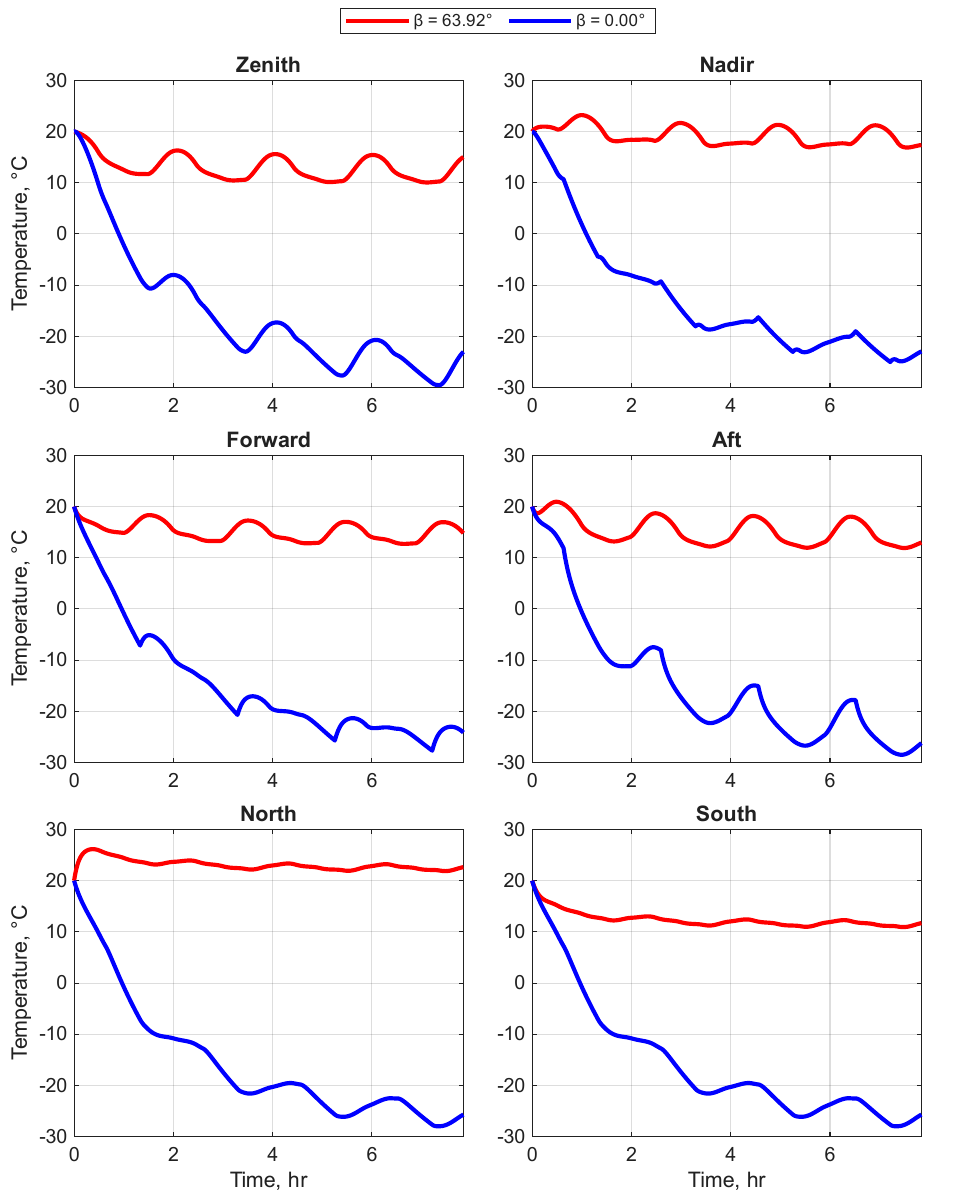}
\caption{Temperatures of an example CubeSat orbiting Mars output in SATMO's specific analysis mode (at distinct beta angles) over four orbital periods (28,237 s). Temperature peaks correspond to sunlit segments of the orbit, while troughs indicate eclipse periods. Each surface exhibits a unique temperature profile due to asymmetric environmental heat loading. A quasi-steady-state temperature response is beginning to form and is expected to stabilize after several more orbits. Simulation parameters are provided in Table \ref{tab: SATMO example simulation props} and Table \ref{tab: SATMO example satellite props}.}
\label{fig: Special Panel Temps} 
\end{figure}

\begin{figure}[H]
\centering
\includegraphics{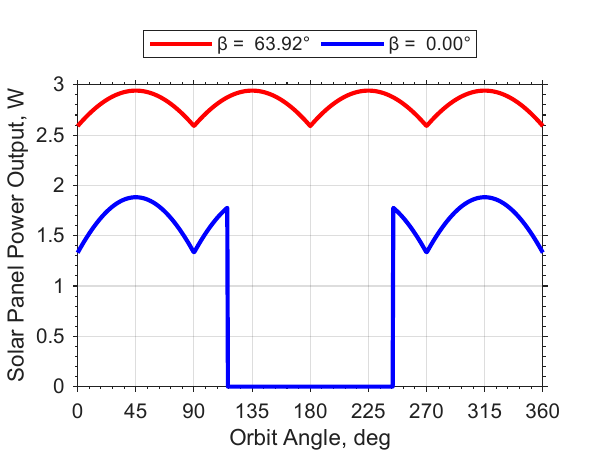}
\caption{Total instantaneous solar panel power output for an example CubeSat orbiting Mars in SATMO's specific analysis mode (at distinct beta angles) over one orbital period (7,059.25 s). Simulation parameters are provided in Table \ref{tab: SATMO example simulation props} and Table \ref{tab: SATMO example satellite props}.}
\label{fig: Special Solar Panel Power} 
\end{figure}

The specific analysis outputs in the hot and cold case for each satellite surface include: (1) the net absorbed environmental heat fluxes from individual direct solar, albedo, and IR planetshine contributions in Fig.~\ref{fig: Special Absorbed Fluxes (Individual)}; (2) the net absorbed environmental heat fluxes from the sum of direct solar, albedo, and IR planetshine contributions in Fig.~\ref{fig: Special Absorbed Fluxes (Total)}; (3) the temperature profiles in Fig.~\ref{fig: Special Panel Temps}; and (4) the minimum and maximum temperatures experienced in Table~\ref{tab: Mars Example Min Max Temp}. The specific analysis also presents the total instantaneous electrical power output by the solar panels over a full orbit, shown in Fig.~\ref{fig: Special Solar Panel Power}.

Figure~\ref{fig: Special Absorbed Fluxes (Individual)} highlights the key elements SATMO's thermal model presented in Section \ref{sec: Thermal Model Development}. The environmental heat flux on each surface depends on its view factor to the primary body, its orientation relative to each heating source, and/or the satellite's angular position along its orbit. Notably, the Zenith surface receives only direct solar radiation, as it has no view to the primary body. For all other surfaces, IR flux remains constant along the entire orbit. The Nadir surface experiences the largest IR and albedo fluxes due to its maximum view of the primary body, while the Forward, Aft, North, and South surfaces exhibit identical IR and albedo flux patterns. In the hot case, the North surface receives a constant solar flux because $\beta = \beta^*$, whereas in the cold case it receives none. The South surface receives no solar flux in either case. Figure~\ref{fig: Special Absorbed Fluxes (Total)} shows the combined environmental heating effects, where, in the hot case, the North surface absorbs the highest total heat flux over a full orbit, while in the cold case, the Nadir surface absorbs the most.

The team can see from Fig.~\ref{fig: Special Panel Temps} and Table~\ref{tab: Mars Example Min Max Temp} that the maximum surface temperatures in the hot case do not deviate significantly from their initial room-temperature values. In the cold case, however, the satellite surfaces can reach temperatures near -30$^\circ$C. Whether this poses a risk depends on the qualification limits of the selected system components. Typical safe operational and survival temperature ranges for general spacecraft components are provided in \cite{gilmore_spacecraft_1999}. To prevent excessively low temperatures, the team could consider applying alternative coatings to their satellite with a higher $\alpha / \varepsilon$ ratio or implementing heater systems and control strategies to maintain components within safe operating limits. Using SATMO's six-surface model, teams can iteratively model their hardware and mission design, re-run simulations, and evaluate the thermal impacts of design changes. This preliminary modeling approach facilitates rapid trade studies and design iteration prior to using higher-fidelity thermal modeling and analysis software.

\begin{table}[H]
\caption{\label{tab: Mars Example Min Max Temp} Minimum and Maximum surface temperatures experienced for an example CubeSat orbiting Mars in SATMO's specific analysis mode (at distinct beta angles) over four orbital periods (28,237 s). Simulation parameters are provided in Table \ref{tab: SATMO example simulation props} and Table \ref{tab: SATMO example satellite props}.}
\centering
\begin{tabular}{lcccccc}
\hline
                                        & Zenith & Nadir & Forward & Aft & North & South \\\hline
\multicolumn{7}{c}{Hot Case, $\beta = 63.92^\circ$}\\
Minimum Temperature, $^\circ \text{C}$  & 10.05 & 16.88 & 12.71 & 11.90 & 20.00 & 10.93 \\
Maximum Temperature, $^\circ \text{C}$  & 20.00 & 23.21 & 20.00 & 20.93 & 26.15 & 20.00 \\
\multicolumn{7}{c}{Cold Case, $\beta = 0.00^\circ$}\\
Minimum Temperature, $^\circ \text{C}$  & -29.53 & -25.02 & -27.68 & -28.49 & -27.98 & -27.98 \\
Maximum Temperature, $^\circ \text{C}$  & 20.01 & 20.00 & 20.00 & 20.00 & 20.00 & 20.00 \\
\hline
\end{tabular}
\end{table}

\section{Summary}
This paper introduces SATMO, a six-node orbital thermal analysis model designed to support rapid space mission planning and the development of thermal management strategies for CubeSats orbiting Earth and other Solar System bodies. The modeling approach builds upon features of previous orbital thermal models and was outlined in detail. Surface temperatures and absorbed environmental heat fluxes calculated by SATMO were then validated and compared to Thermal Desktop (TD) simulations for an arbitrary 1U CubeSat around Venus, Earth, and Mars. The largest discrepancies in absorbed environmental heat fluxes calculated by SATMO and TD were associated with the albedo heating contributions, reaching $\approx50$ \si[per-mode=symbol]{\watt\per\m\squared} during Test Case 7. The largest discrepancies in satellite surface temperatures calculated by SATMO and TD were $\approx1.2^\circ$C, occurring during Test Case 5. An example use case involving a 1U Mars CubeSat was presented to further demonstrate SATMO's functionality.

Future extensions to SATMO include: (1) validating the tool for satellites orbiting the Moon and the remaining planets in the Solar System; (2) enabling elliptical orbit modeling; (3) incorporating radiation exchange between internal satellite surfaces and components; (4) supporting non-nadir-pointing configurations (e.g., spinning and slewing); (5) removing the uniform primary body surface illumination assumption; (6) simulating thermal effects on spacecraft with deployed solar arrays; and (7) validating this work other CubeSat sizes after implementing the aforementioned improvements. By increasing access to user-friendly thermal modeling tools, SATMO addresses an important need in the early stages of spacecraft mission planning. As research groups pursue more interplanetary SmallSat initiatives, SATMO is positioned to become a valuable addition to their thermal analysis toolkit.


\section*{Acknowledgments}
This material is based upon work supported by the National Science Foundation (NSF) Graduate Research Fellowship Program under Grant No. DGE-2039655. The research was carried out at the Jet Propulsion Laboratory, California Institute of Technology, under a contract with the National Aeronautics and Space Administration (80NM0018D0004). Artificial Intelligence (AI) tools (OpenAI ChatGPT) were occasionally used to assist with improving readability, grammar, and language in the manuscript text. No AI tools were used for data analysis, figure generation, or interpretation of results. Any opinions, findings, and conclusions or recommendations expressed in this material are those of the author(s) and do not necessarily reflect the views of NSF or JPL.
\bibliography{MITSM}
\end{document}